\documentclass[10pt]{article}
\usepackage{graphicx}
\usepackage{amsmath}
\usepackage{amssymb}
\usepackage{balance}
\usepackage{bm}
\usepackage{caption2}
\usepackage{indentfirst}

\begin{document}
\bibliographystyle{prsty}
\begin{center}
{\large {\bf \sc{  Scalar hidden-charm tetraquark states with QCD sum rules   }}} \\[2mm]
Zun-Yan Di$^{1,2}$, Zhi-Gang Wang$^{1}$\footnote{E-mail: zgwang@aliyun.com. }, Jun-Xia Zhang$^{1,2}$, Guo-Liang Yu$^{1}$    \\
$^{1}$ Department of Physics, North China Electric Power University, Baoding 071003, P. R. China \\
$^{2}$ School of Nuclear Science and Engineering, North China Electric Power University, Beijing 102206, P. R. China
\end{center}

\begin{abstract}
In this article, we study the masses and pole residues of the pseudoscalar-diquark-pseudoscalar-antidiquark type and vector-diquark-vector-antidiquark type scalar hidden-charm $cu\bar{c}\bar{d}$ ($cu\bar{c}\bar{s}$) tetraquark states with QCD sum rules by taking into account the contributions of the vacuum condensates up to dimension-10 in the operator product expansion. The predicted  masses  can be confronted with the experimental data in the future. Possible decays of those tetraquark states are also discussed.
\end{abstract}



\section{Introduction}

In recent years, a  number of charged charmonium-like (bottomonium-like) exotic states have been observed, such as the $Z_c(3900)$, $Z_c(4020)$, $Z_c(4050)$, $Z_c(4055)$, $Z_c(4200)$, $Z_c(4240)$, $Z_c(4250)$, $Z(4430)$, $Z_b(10610)$ and $Z_b(10650)$ \cite{PDG}. If those  charged charmonium-like (bottomonium-like) states are resonances indeed, their quark constituents must be $c\bar{c}u\bar{d}$ or $c\bar{c}d\bar{u}$ ($b\bar{b}u\bar{d}$ or $b\bar{b}d\bar{u}$), irrespective of the diquark-antidiquark type or meson-meson type substructures.

Those  exotic states cannot be accommodated within the naive quark model, and represent a new facet of QCD and provide a new
opportunity for a deeper understanding of the non-perturbative QCD.  The QCD sum rules method is a powerful tool in studying the
hidden-charm (bottom) tetraquark or molecular states  and hidden-charm pentaquark states \cite{No-formular,WangHuang-Tetraquark,WangHuang-molecule}.
In the QCD sum rules, the operator product expansion is used to expand the time-ordered currents into a series of quark and gluon
condensates which parameterize the non-perturbative properties of the QCD vacuum. Based on the quark-hadron duality, we can obtain copious
information about the hadronic parameters at the phenomenological side \cite{SVZ79,PRT85}.

The diquarks $\varepsilon^{ijk}q^{T}_j C\Gamma q^{\prime}_k$ have  five  structures  in Dirac spinor space, where $C\Gamma=C\gamma_5$, $C$, $C\gamma_\mu \gamma_5$,  $C\gamma_\mu $ and $C\sigma_{\mu\nu}$ for the scalar, pseudoscalar, vector, axialvector  and  tensor diquarks, respectively. In this expression, $q_j$ denotes the quark field; $i$, $j$ and $k$ are color indexes; $C$ is the charge conjugation matrix; and the superscript $T$ denotes the transpose of the Dirac indexes.
The attractive interactions of one-gluon exchange  favor  formation of
the diquarks in  color antitriplet, flavor
antitriplet and spin singlet \cite{One-gluon},
 while the favored configurations are the scalar ($C\gamma_5$) and axialvector ($C\gamma_\mu$) diquark states \cite{WangDiquark,WangLDiquark}.
 We can construct the  diquark-antidiquark type hidden charm tetraquark states \cite{WangScalarT},
\begin{eqnarray}
&&C\gamma_5 \otimes \gamma_5C\, , \nonumber\\
&&C\gamma_\mu \otimes \gamma^\mu C\, ,
\end{eqnarray}
   to study the lowest scalar tetraquark states, or construct the
\begin{eqnarray}
&&C \otimes C\, , \nonumber\\
&&C\gamma_\mu \gamma_5 \otimes \gamma_5\gamma^\mu C\, ,
\end{eqnarray}
to study the scalar tetraquark states having larger masses.

 In this article,   we construct $C\otimes C$ and $C\gamma_{\mu}\gamma_{5}\otimes \gamma_{5}\gamma^{\mu}C$  type   currents  to explore the charged scalar
hidden-charm tetraquark states by calculating the contributions of
the vacuum condensates up to dimension-10 in a consistent way.

This article is organized as follows: in section 2, we derive the QCD sum rules to extract the masses and pole residues of the
charged scalar $cu\bar{c}\bar{d}$ ($cu\bar{c}\bar{s}$) tetraquark states; in section 3, we present the numerical
results and discussions; section 4 is reserved for conclusion.

\section{QCD sum rules for the scalar hidden-charm tetraquark states}
In QCD sum rules, we consider the two-point correlation functions $\Pi_{1,2,3,4}(p)$,
\begin{eqnarray}\label{correlation function}
\Pi_{1,2,3,4}(p)&=&i\int d^{4} x e^{ip\cdot x} \langle0|T\left\{J_{1,2,3,4}(x) J_{1,2,3,4}^{\dag}(0)\right\}|0\rangle\ ,
\end{eqnarray}
where the $J_{1,2,3,4}(x)$ are the interpolating currents with the same
quantum numbers as the tetraquark states we want to study. Those  currents are constructed in the diquark model
and can be  written   as,
\begin{eqnarray}
J_{1}(x)&=&\varepsilon^{i j k}\varepsilon^{i m n} u^{j T} (x)Cc^{k} \bar{d}^{m}(x) C \bar{c}^{n T}(x)\ ,  \\
J_{2}(x)&=&\varepsilon^{i j k}\varepsilon^{i m n} u^{j T} (x)C\gamma_{\mu}\gamma_{5}c^{k} \bar{d}^{m}(x) \gamma_{5}\gamma^{\mu}C \bar{c}^{n T}(x)\ , \\
J_{3}(x)&=&\varepsilon^{i j k}\varepsilon^{i m n} u^{j T} (x)Cc^{k} \bar{s}^{m}(x) C \bar{c}^{n T}(x)\ ,  \\
J_{4}(x)&=&\varepsilon^{i j k}\varepsilon^{i m n} u^{j T} (x)C\gamma_{\mu}\gamma_{5}c^{k} \bar{s}^{m}(x) \gamma_{5}\gamma^{\mu}C \bar{c}^{n T}(x)\ .
\end{eqnarray}

On  the phenomenological side, we insert a complete set of intermediate hadronic states with the same quantum numbers as the
current operators $J_{1,2,3,4}(x)$ into the correlation functions $\Pi_{1,2,3,4}(p)$ to obtain the hadronic representations \cite{SVZ79,PRT85}. After isolating the ground state contributions of the scalar
tetraquark states, we get the following results,
\begin{eqnarray}
\Pi_{1,2,3,4}(p)&=&\frac{\lambda^{2}_{Z_{1,2,3,4}}}{M^{2}_{Z_{1,2,3,4}}-p^{2}}+\cdots \ ,
\end{eqnarray}
where the pole residues $\lambda_{Z_{1,2,3,4}}$ are defined by $\langle0|J_{1,2,3,4}(0)|Z_{1,2,3,4}(p)\rangle=\lambda_{Z_{1,2,3,4}}$.

At the quark level, the two-point correlation functions $\Pi_{1,2,3,4}(p)$  can be evaluated via the operator product expansion  method. We   contract the $u$, $d$, $c$ and $s$ quark fields with the wick theorem and obtain the following results:
\begin{eqnarray}
\Pi_{1}(p)&=&i\varepsilon^{i j k}\varepsilon^{i m n}\varepsilon^{i' j' k'}\varepsilon^{i' m' n'}\int d^{4}x e^{ip\cdot x} \nonumber\\
&&{\rm Tr}\left[C^{k k'}(x)CU^{j j' T}(x)C\right]{\rm Tr}\left[C^{n' n}(-x)CD^{m' m T}(-x)C\right]\ ,\\
\Pi_{2}(p)&=&i\varepsilon^{i j k}\varepsilon^{i m n}\varepsilon^{i' j' k'}\varepsilon^{i' m' n'}\int d^{4}x e^{ip\cdot x} \nonumber\\
&&{\rm Tr}\left[\gamma_{\mu}\gamma_{5}C^{k k'}(x)\gamma_{5}\gamma_{\nu}CU^{j j' T}(x)C\right]{\rm Tr}\left[\gamma_{5}\gamma^{\nu}C^{n' n}(-x)\gamma^{\mu}\gamma_{5}CD^{m' m T}(-x)C\right]\ ,\\
\Pi_{3}(p)&=&i\varepsilon^{i j k}\varepsilon^{i m n}\varepsilon^{i' j' k'}\varepsilon^{i' m' n'}\int d^{4}x e^{ip\cdot x} \nonumber\\
&&{\rm Tr}\left[C^{k k'}(x)CU^{j j' T}(x)C\right]{\rm Tr}\left[C^{n' n}(-x)CS^{m' m T}(-x)C\right]\ ,\\
\Pi_{4}(p)&=&i\varepsilon^{i j k}\varepsilon^{i m n}\varepsilon^{i' j' k'}\varepsilon^{i' m' n'}\int d^{4}x e^{ip\cdot x} \nonumber\\
&&{\rm Tr}\left[\gamma_{\mu}\gamma_{5}C^{k k'}(x)\gamma_{5}\gamma_{\nu}CU^{j j' T}(x)C\right]{\rm Tr}\left[\gamma_{5}\gamma^{\nu}C^{n' n}(-x)\gamma^{\mu}\gamma_{5}CS^{m' m T}(-x)C\right]\ ,
\end{eqnarray}
where the $U_{i j}(x)$, $D_{i j}(x)$, $S_{i j}(x)$ and $C_{i j}(x)$
are the full $u$, $d$, $s$ and $c$ quark propagators,  respectively.
For simplicity, the $U_{i j}(x)$ and $D_{i j}(x)$ can be written as
$P_{i j}(x)$,
\begin{eqnarray}
P_{i j}(x)&=&\frac{i\delta_{i j}x\!\!\!/}{2\pi^{2}x^{4}}-\frac{\delta_{i j}\langle\bar{q}q\rangle}{12}-\frac{\delta_{i j}x^{2}\langle\bar{q}g_{s}\sigma Gq\rangle}{192}-\frac{\delta_{i j}x^{2}x\!\!\!/g_{s}^{2}\langle\bar{q}q\rangle^{2}}{7776}-\frac{i g_{s}G_{\alpha\beta}^{n}t_{i j}^{n}(x\!\!\!/\sigma^{\alpha\beta}+\sigma^{\alpha\beta}x\!\!\!/)}{32\pi^{2}x^{2}} \nonumber\\
&&-\frac{\delta_{i j}x^{4}\langle\bar{q}q\rangle\langle GG\rangle}{27648}-\frac{1}{8}\langle\bar{q}_{j}\sigma^{\alpha\beta}q_{i}\rangle\sigma_{\alpha\beta}-\frac{1}{4}\langle\bar{q}_{j}\gamma_{\mu}q_{i}\rangle\gamma^{\mu}+\cdots \ ,\label{propagator-u,d}\\
S_{i j}(x)&=&\frac{i\delta_{i j}x\!\!\!/}{2\pi^{2}x^{4}}-\frac{\delta_{i j}m_{s}}{4\pi^{2}x^{2}}-\frac{\delta_{i j}\langle\bar{s}s\rangle}{12}+\frac{i\delta_{i j}x\!\!\!/m_{s}\langle\bar{s}s\rangle}{48}-\frac{\delta_{i j}x^{2}\langle\bar{s}g_{s}\sigma Gs\rangle}{192} \nonumber\\
&&+\frac{i\delta_{i j}x^{2}x\!\!\!/m_{s}\langle\bar{s}g_{s}\sigma Gs\rangle}{1152}-\frac{\delta_{i j}x^{2}x\!\!\!/g_{s}^{2}\langle\bar{s}s\rangle^{2}}{7776}-\frac{i g_{s}G_{\alpha\beta}^{n}t_{i j}^{n}(x\!\!\!/\sigma^{\alpha\beta}+\sigma^{\alpha\beta}x\!\!\!/)}{32\pi^{2}x^{2}} \nonumber\\
&&-\frac{\delta_{i j}x^{4}\langle\bar{s}s\rangle\langle GG\rangle}{27648}-\frac{1}{8}\langle\bar{s}_{j}\sigma^{\alpha\beta}s_{i}\rangle\sigma_{\alpha\beta}-\frac{1}{4}\langle\bar{s}_{j}\gamma_{\mu}s_{i}\rangle\gamma^{\mu}+\cdots \ ,\label{propagator-s}\\
C_{i j}(x)&=&\frac{i}{(2\pi)^4}\int d^4 ke^{-ik\cdot x}\bigg\{\frac{k\!\!\!/ +m_{c}}{k^{2}-m_{c}^{2}}\delta_{i j}-g_{s}t_{i j}^{n}G_{\alpha\beta}^{n}\frac{(k\!\!\!/+m_{c})\sigma^{\alpha\beta}+\sigma^{\alpha\beta}(k\!\!\!/+m_{c})}{4(k^{2}-m_{c}^{2})^{2}} \nonumber\\
&&+\frac{g_{s}t_{i j}^{n}D_{\alpha}G_{\beta\lambda}^{n}(f^{\lambda\alpha\beta}+f^{\lambda\beta\alpha})}{3(k^{2}-m_{c}^{2})^{4}} \nonumber\\
&&-\frac{g_{s}^{2}(t^{n}t^{m})_{i j}G_{\alpha\beta}^{n}G_{\mu\nu}^{n}(f^{\alpha\beta\mu\nu}+f^{\alpha\mu\beta\nu}+f^{\alpha\mu\nu\beta})}{4(k^{2}-m_{c}^2)^{5}}+\cdots\bigg\} \ ,
\end{eqnarray}
\begin{eqnarray}
f^{\lambda\alpha\beta}&=&(k\!\!\!/+m_{c})\gamma^{\lambda}(k\!\!\!/+m_{c})\gamma^{\alpha}(k\!\!\!/+m_{c})\gamma^{\beta}(k\!\!\!/+m_{c})\ ,\nonumber\\
f^{\alpha\beta\mu\nu}&=&(k\!\!\!/+m_{c})\gamma^{\alpha}(k\!\!\!/+m_{c})\gamma^{\beta}(k\!\!\!/+m_{c})\gamma^{\mu}(k\!\!\!/+m_{c})\gamma^{\nu}(k\!\!\!/+m_{c})\ ,
\end{eqnarray}
and $t^{n}=\frac{\lambda^{n}}{2}$, the $\lambda^{n}$ is the
Gell-Mann matrix, $D_{\alpha}=\partial
_\alpha-ig_{s}G_{\alpha}^{n}t^{n}$ \cite{PRT85}. In
Eqs.\eqref{propagator-u,d}--\eqref{propagator-s}, we retain the terms
$\langle\bar{q}_{j}\sigma_{\mu\nu}q_{i}\rangle$, $\langle\bar{s}_{j}\sigma_{\mu\nu}s_{i}\rangle$,
$\langle\bar{q}_{j}\gamma_{\mu}q_{i}\rangle$ and $\langle\bar{s}_{j}\gamma_{\mu}s_{i}\rangle$ originate from the
Fierz re-arrangement of the $\langle q_{i}\bar{q}_{j}\rangle$, $\langle s_{i}\bar{s}_{j}\rangle$ to absorb the gluons emitted from
the heavy quark lines so as to extract the mixed condensates and four-quark condensates $\langle\bar{q}g_{s}\sigma Gq\rangle$,
$\langle\bar{s}g_{s}\sigma Gs\rangle$, $g_{s}^{2}\langle\bar{q}q\rangle^{2}$ and
$g_{s}^{2}\langle\bar{s}s\rangle^{2}$, respectively. We compute  the integrals both in the coordinate and momentum spaces by taking into
account the contributions of the vacuum condensates up to dimension-10. Then, we obtain the QCD spectral densities from the
imaginary parts  of the correlations.

After getting the QCD spectral densities, we take the quark-hadron duality below
the continuum thresholds $s_{0}$ and perform the Borel transformation with respect to the variable $P^{2}=-p^{2}$ to obtain
the  QCD sum rules,
\begin{eqnarray}\label{PoleResidue}
\lambda_{Z_{1,2,3,4}}^{2}\exp\left(-\frac{M_{Z_{1,2,3,4}}^{2}}{T^{2}}\right)&=&\int_{4m_{c}^{2}}^{s_{0}^{1,2,3,4}}ds\rho^{1,2,3,4}\left(s\right)\exp\left(-\frac{s}{T^{2}}\right)\ ,
\end{eqnarray}
where
\begin{eqnarray}
\rho^{1,2,3,4}\left(s\right)&=& \rho_{0}^{1,2,3,4}\left(s\right)+\rho_{3}^{1,2,3,4}\left(s\right)+\rho_{4}^{1,2,3,4}\left(s\right)+\rho_{5}^{1,2,3,4}\left(s\right)+\rho_{6}^{1,2,3,4}\left(s\right)+\rho_{7}^{1,2,3,4}\left(s\right) \nonumber\\
&&+\rho_{8}^{1,2,3,4}\left(s\right)+\rho_{10}^{1,2,3,4}\left(s\right)\ ,
\end{eqnarray}
the explicit expressions of the spectral densities
$\rho^{1,2,3,4}(s)$ are given in the appendix. The subscripts 0, 3,
4, 5, 6, 7, 8, 10 denote the dimensions of the vacuum condensates.
We take into account  the vacuum condensates which are vacuum expections of the
operators of the orders $O(\alpha_{s}^{k})$ with $k\leq1$
consistently.

We derive Eq.\eqref{PoleResidue} with respect to $\frac{1}{T^{2}}$, then eliminate the pole residues $\lambda_{Z_{1,2,3,4}}$, and obtain the
expressions for the masses of the  scalar tetraquark states,
\begin{eqnarray}\label{mass}
M_{Z_{1,2,3,4}}^{2}&=&\frac{\int_{4m_{c}^{2}}^{s_{0}^{1,2,3,4}}ds\frac{d}{d\left(-1/T^{2}\right)}\rho^{1,2,3,4}\left(s\right)\exp\left(-\frac{s}{T^{2}}\right)} {\int_{4m_{c}^{2}}^{s_{0}^{1,2,3,4}}ds\rho^{1,2,3,4}\left(s\right)\exp\left(-\frac{s}{T^{2}}\right)}\ .
\end{eqnarray}
Once the masses are obtained, we can take them as input parameters
and obtain the pole residues from the QCD sum sules in
Eq.\eqref{PoleResidue}.

\section{Numerical results and discussions}

In this section, we perform the numerical analysis, and choose the reasonable QCD parameters for the quark masses and vacuum
condensates. The vacuum condensates are taken to be the standard
values $\langle\bar{q}q\rangle=-(0.24\pm0.01\,\text{GeV})^{3}$, $\langle\bar{s}s\rangle=(0.8\pm0.1)\langle\bar{q}q\rangle$,
$\langle\bar{q}g_{s}\sigma Gq\rangle=m_{0}^{2}\langle\bar{q}q\rangle$,  $\langle\bar{s}g_{s}\sigma Gs\rangle=m_{0}^{2}\langle\bar{s}s\rangle$,
$m_{0}^{2}=(0.8\pm0.1)\,\text{GeV}^{2}$, $\langle\frac{\alpha_{s}GG}{\pi}\rangle=(0.33\,\text{GeV})^{4}$ at
the energy scale  $\mu=1\,\text{GeV}$ \cite{SVZ79,PRT85,ColangeloReview}. The quark
condensates and mixed quark condensates evolve with the renormalization group equation,
$\langle\bar{q}q\rangle(\mu)=\langle\bar{q}q\rangle(Q)[\frac{\alpha_{s}(Q)}{\alpha_{s}(\mu)}]^{\frac{4}{9}}$,
$\langle\bar{s}s\rangle(\mu)=\langle\bar{s}s\rangle(Q)[\frac{\alpha_{s}(Q)}{\alpha_{s}(\mu)}]^{\frac{4}{9}}$,
$\langle\bar{q}g_{s}\sigma Gq\rangle(\mu)=\langle\bar{q}g_{s}\sigma
Gq\rangle(Q)[\frac{\alpha_{s}(Q)}{\alpha_{s}(\mu)}]^{\frac{2}{27}}$
and $\langle\bar{s}g_{s}\sigma
Gs\rangle(\mu)=\langle\bar{s}g_{s}\sigma
Gs\rangle(Q)[\frac{\alpha_{s}(Q)}{\alpha_{s}(\mu)}]^{\frac{2}{27}}$.
In addition, we take the values
$m_{u}(\mu=1\,\text{GeV})=m_{d}(\mu=1\,\text{GeV})=m_{q}(\mu=1\,\text{GeV})=0.006\,\text{GeV}$
from the Gell-Mann-Oakes-Renner relation, and choose the
$\overline{MS}$ mass $m_{c}(m_{c})=(1.275\pm0.025)\,\text{GeV}$ and
$m_{s}(\mu=2\,\text{GeV})=(0.095\pm0.005)\,\text{GeV}$ from the
Particle Data Group \cite{PDG}, and take into account the
energy-scale dependence of the $\overline{MS}$ masses from the
renormalization group equation,
\begin{eqnarray}
m_{q}(\mu)&=&m_{q}\left(1\,\text{GeV}\right)\bigg[\frac{\alpha_{s}(\mu)}{\alpha_{s}(1\,\text{GeV})}\bigg]^{\frac{4}{9}}\ ,\nonumber\\
m_{s}(\mu)&=&m_{s}\left(2\,\text{GeV}\right)\bigg[\frac{\alpha_{s}(\mu)}{\alpha_{s}(2\,\text{GeV})}\bigg]^{\frac{4}{9}}\ ,\nonumber\\
m_{c}(\mu)&=&m_{c}\left(m_{c}\right)\bigg[\frac{\alpha_{s}(\mu)}{\alpha_{s}(m_{c})}\bigg]^{\frac{12}{25}}\ ,\nonumber\\
\alpha_{s}(\mu)&=&\frac{1}{b_{0}t}\left[1-\frac{b_{1}}{b_{0}^{2}}\frac{\log t}{t}+\frac{b_{1}^{2}\left(\log^{2}t-\log t-1\right)+b_{0}b_{2}}{b_{0}^{4}t^{2}}\right]\ ,
\end{eqnarray}
where $t=\log \frac{\mu^{2}}{\Lambda^{2}}$, $b_{0}=\frac{33-2n_{f}}{12\pi}$, $b_{1}=\frac{153-19n_{f}}{24\pi^{2}}$, $b_{2}=\frac{2857-\frac{5033}{9}n_{f}+\frac{325}{27}n_{f}^{2}}{128\pi^{3}}$, $\Lambda=213\,\text{MeV}$, $296\,\text{MeV}$ and $339\,\text{MeV}$ for the flavors $n_{f}=5$, $4$ and $3$, respectively \cite{PDG}.

\begin{figure}[htp]
\centering
\includegraphics[totalheight=5cm,width=7cm]{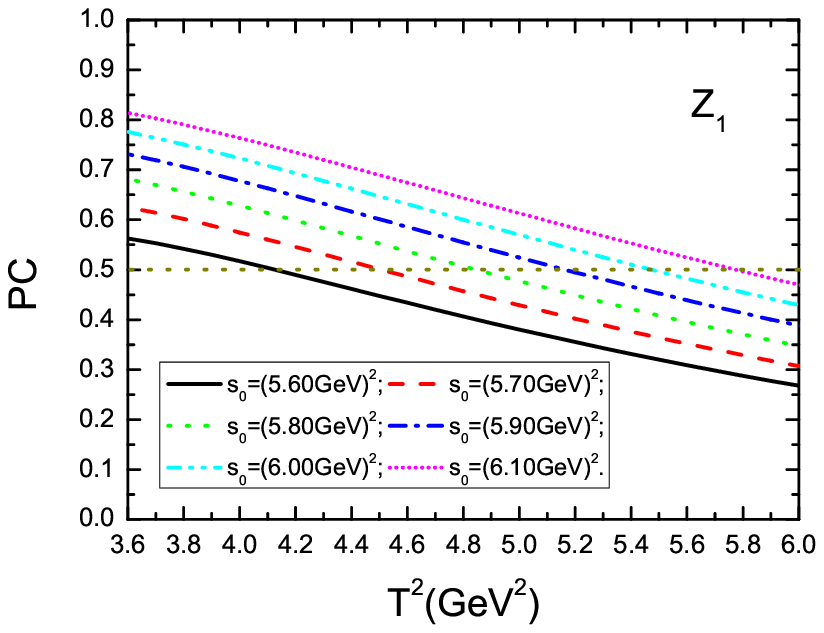}
\includegraphics[totalheight=5cm,width=7cm]{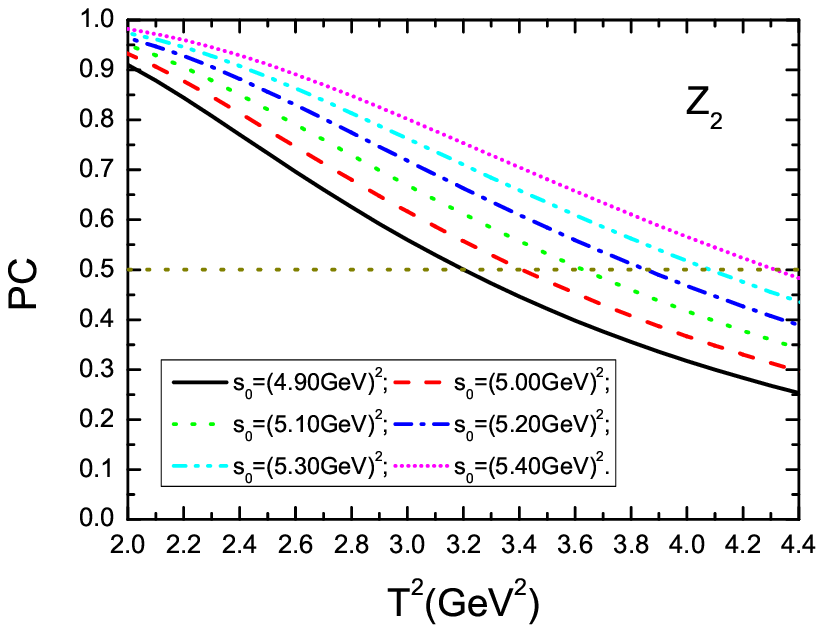}
\includegraphics[totalheight=5cm,width=7cm]{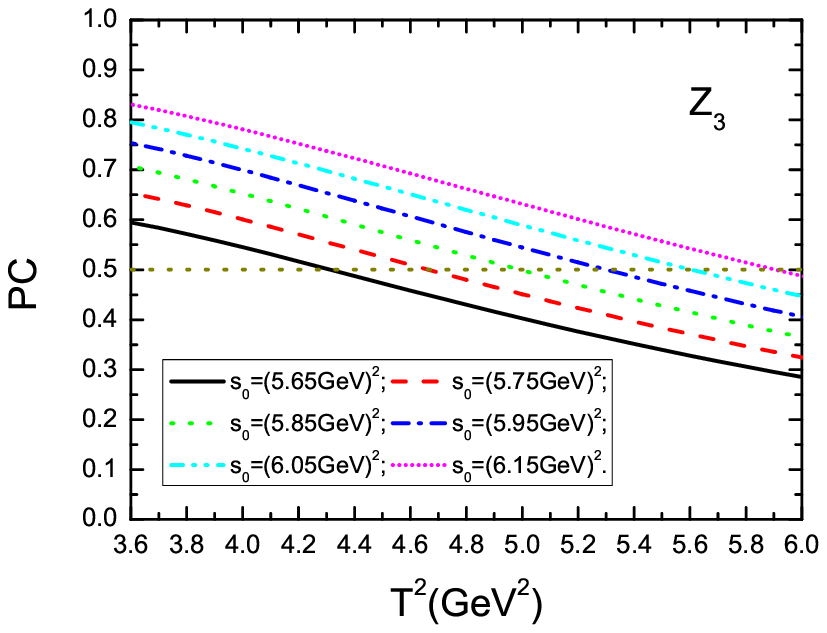}
\includegraphics[totalheight=5cm,width=7cm]{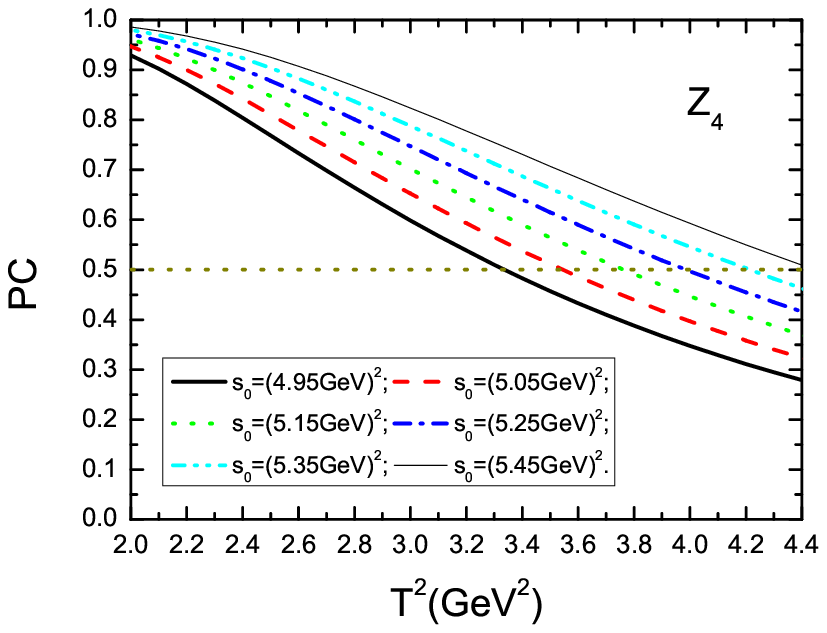}
\caption{The pole contributions with variations of the Borel parameters $T^{2}$ and threshold parameters $s_{0}$.}\label{fig:fig1}
\end{figure}

The energy scale $\mu$ is considered as a variable. In Refs.\cite{WangHuang-Tetraquark,WangHuang-molecule},   the energy
scale dependence of the QCD sum rules for the hidden-charm tetraquark states and molecular states is studied  in details
for the first time, and   an energy scale formula,
\begin{eqnarray}\label{Escalar}
\mu&=&\sqrt{M_{X/Y/Z}^{2}-\left(2{\mathbb{M}}_{c}\right)^{2}}\ ,
\end{eqnarray}
 with the   effective $c$-quark mass ${\mathbb{M}}_{c}$ is suggested. The formula works well for the $X(3872)$, $Z_c(3885/3900)$, $X^*(3860)$, $Y(3915)$, $Z_c(4020/4025)$,   $Z(4430)$, $X(4500)$, $Y(4630/4660)$, $X(4700)$ in the scenario of tetraquark states.  In this article, we take the updated value of the effective $c$-quark mass ${\mathbb{M}}_c=1.82\,\rm{GeV}$ to determine the energy scales of the QCD spectral densities \cite{WangEPJC4260}.

The mass gaps between the ground states
and the first radial excited states are usually taken as
$(0.4-0.6)\,\text{GeV}$. For examples, the $Z(4430)$ is tentatively
assigned to be the first radial excitation of the $Z_{c}(3900)$
according to the analogous decays, $Z_{c}(3900)^{\pm} \longrightarrow  J/\psi \pi^{\pm}$, $Z(4430)^{\pm} \longrightarrow \psi'\pi^{\pm}$
and the mass differences $M_{Z(4430)}-M_{Z_{c}(3900)}=576\,\text{MeV}$, $M_{\psi'}-M_{J/\psi}=589\,\text{MeV}$ \cite{WangZG-Z4430, Z4430-1405,Nielsen-1430005}; the $X(3915)$ and $X(4500)$ are assigned  to be the ground state and the first radial excited state of the axialvector-diquark-axialvector-antidiquark type scalar $cs\bar{c}\bar{s}$ tetraquark states, respectively, and their mass difference is $M_{X(4500)}-M_{X(3915)}=588\,\text{MeV}$ \cite{WangZG-X3915-X4500}. The relation
\begin{eqnarray}\label{s0-relation}
\sqrt{s_{0}}&=&M_{X/Y/Z}+(0.4-0.6)\,\text{GeV}\ ,
\end{eqnarray}
serves as another constraint on the masses of the hidden-charm tetraquark states. In calculations, we observe that the values  of the
masses $M_Z$ decrease with  increase of the energy scales   $\mu$
from QCD sum rules in Eq.\eqref{mass}. While Eq.\eqref{Escalar} indicates
that the value of the masses  $M_Z$ increase when the energy scales
$\mu$ increase. There exist  optimal energy scales, which lead to  reasonable  masses $M_{Z}$.

We should obey two criteria to choose the Borel parameters $T^{2}$ and threshold parameters $s_{0}$ in  numerical calculations.
The first criterion is the pole dominance on the phenomenological side. The pole
contribution (PC) is defined by,
\begin{eqnarray}
\text{PC}&=&\frac{\int_{4m_{c}^{2}}^{s_{0}^{1,2,3,4}}ds\rho^{1,2,3,4}\left(s\right)\exp\left(-\frac{s}{T^{2}}\right)} {\int_{4m_{c}^{2}}^{\infty}ds\rho^{1,2,3,4}\left(s\right)\exp\left(-\frac{s}{T^{2}}\right)}\ .
\end{eqnarray}
The second criterion is the convergence of the operator product expansion. To judge the
convergence, we calculate the contributions $D_i$ in the
operator product expansion with the formula,
\begin{eqnarray}
D_i&=&\frac{\int_{4m_{c}^{2}}^{s_{0}^{1,2,3,4}}ds\rho_{i}^{1,2,3,4}(s)\exp\left(-\frac{s}{T^{2}}\right)}{\int_{4m_{c}^{2}}^{s_{0}^{1,2,3,4}}ds\rho^{1,2,3,4}\left(s\right)\exp\left(-\frac{s}{T^{2}}\right)}\ ,
\end{eqnarray}
where the index $i$ denotes the dimension of the vacuum condensates.

\begin{figure}[htp]
\centering
\includegraphics[totalheight=5cm,width=7cm]{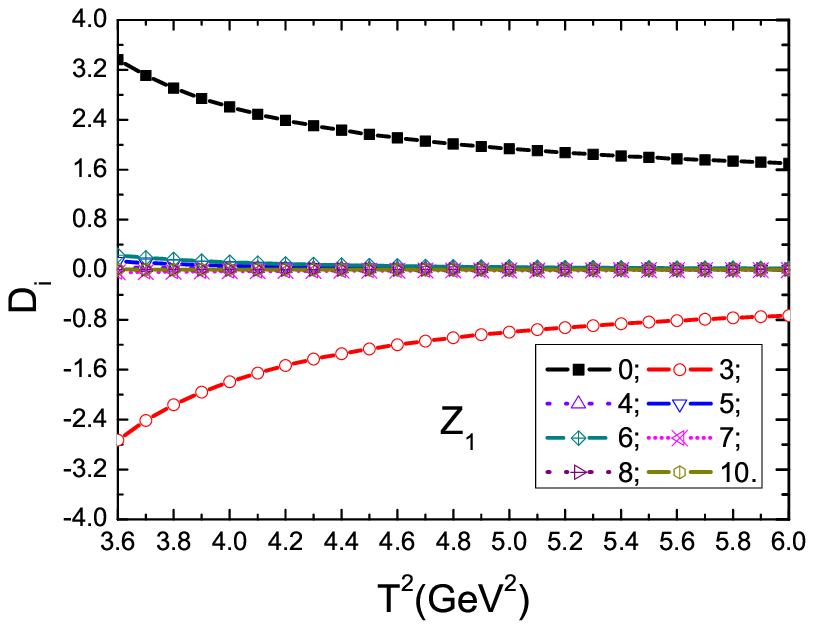}
\includegraphics[totalheight=5cm,width=7cm]{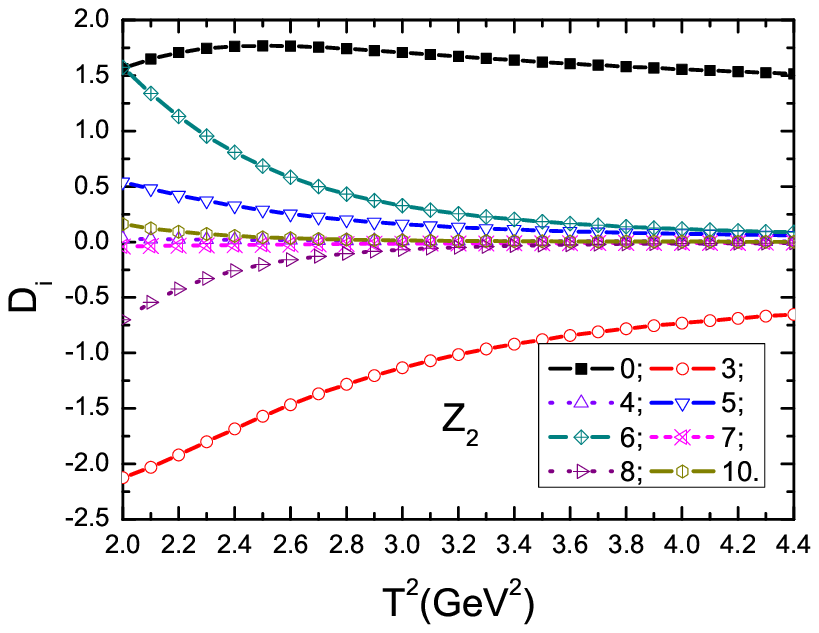}
\includegraphics[totalheight=5cm,width=7cm]{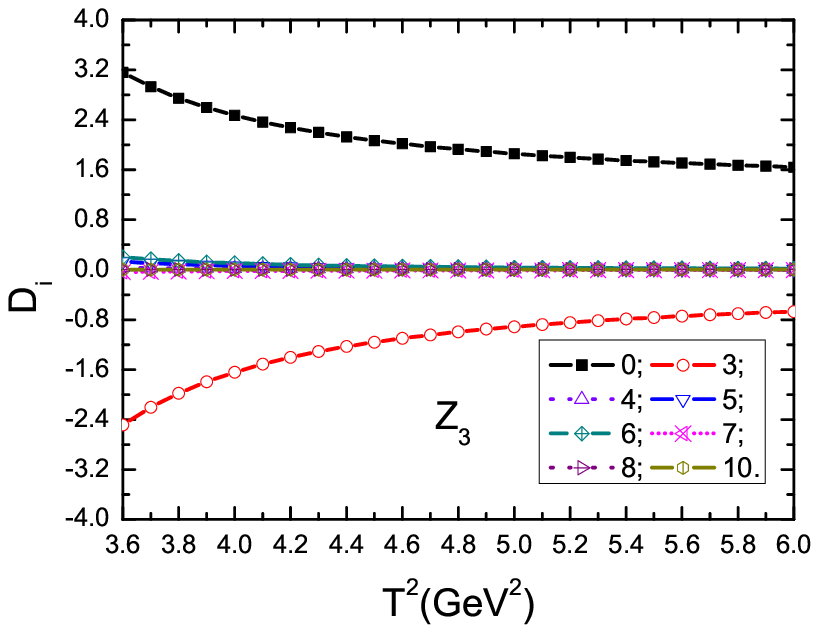}
\includegraphics[totalheight=5cm,width=7cm]{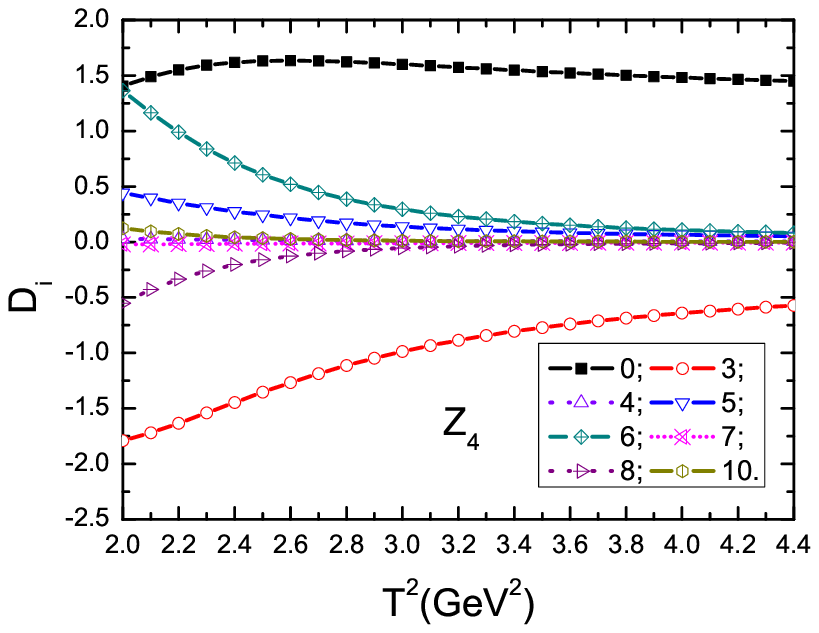}
\caption{The contributions of different terms in the operator product expansion with variations of the Borel parameters $T^{2}$, where the $0$, $3$, $4$, $5$, $6$, $7$, $8$, $10$ denote the dimensions of the vacuum condensates.}\label{fig:fig2}
\end{figure}

In Fig.\ref{fig:fig1}, the contributions of the pole terms are plotted with variations of  the Borel parameters $T^{2}$ for different values of the
threshold parameters $s_{0}$, where the values of energy scales are taken as
$\mu=4.00\,\text{GeV}$, $2.90\,\text{GeV}$, $4.05\,\text{GeV}$
and $2.95\,\text{GeV}$  for the tetraquark states $Z_{1}$, $Z_{2}$, $Z_{3}$, $Z_{4}$, respectively. From the figure, we can see
that the continuum thresholds  $\sqrt{s_{0}}\leq5.70\,\text{GeV}$,
$\leq5.00\,\text{GeV}$,
$\leq5.75\,\text{GeV}$, $\leq5.05\,\text{GeV}$
for the tetraquark states $Z_{1}$, $Z_{2}$, $Z_{3}$, $Z_{4}$ respectively  are too small
to satisfy the pole dominance condition to result in reasonable
Borel windows.

\begin{figure}[htp]
\centering
\includegraphics[totalheight=5cm,width=7cm]{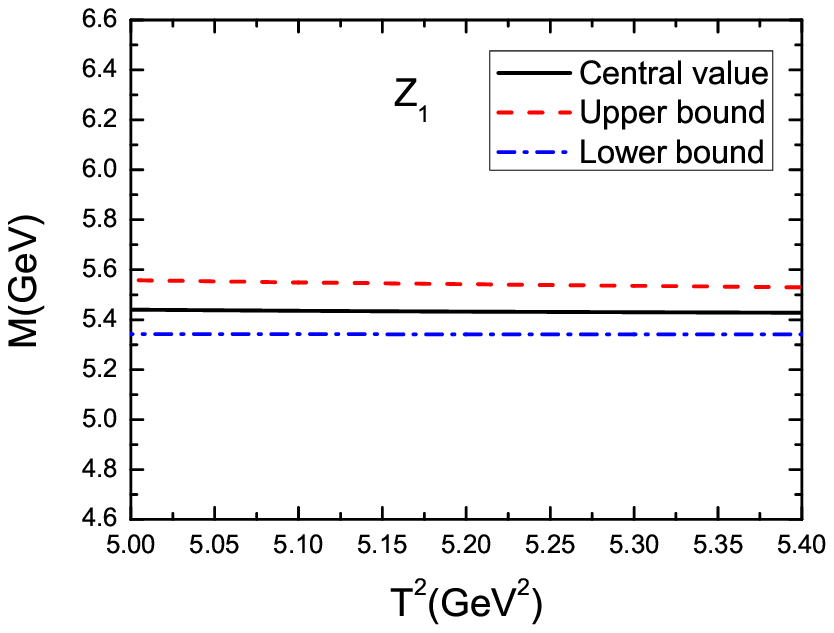}
\includegraphics[totalheight=5cm,width=7cm]{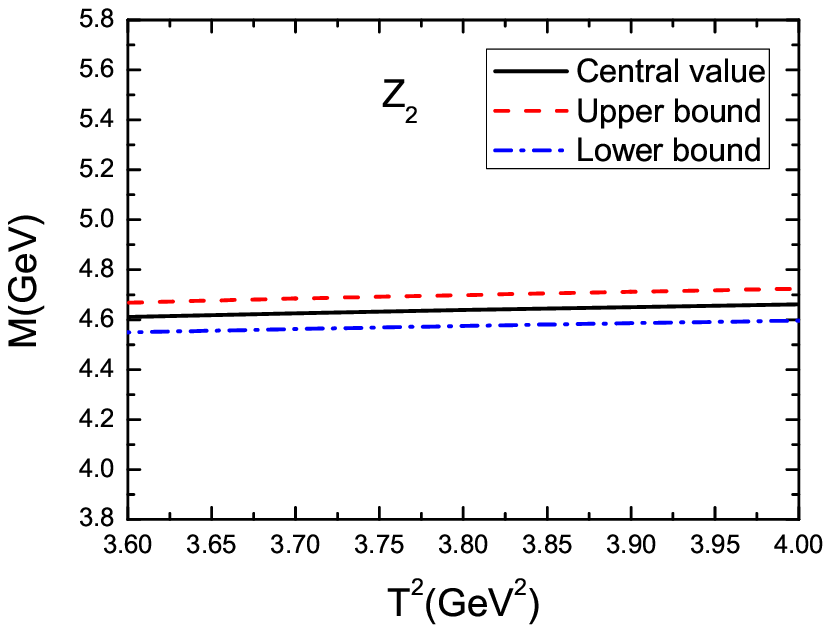}
\includegraphics[totalheight=5cm,width=7cm]{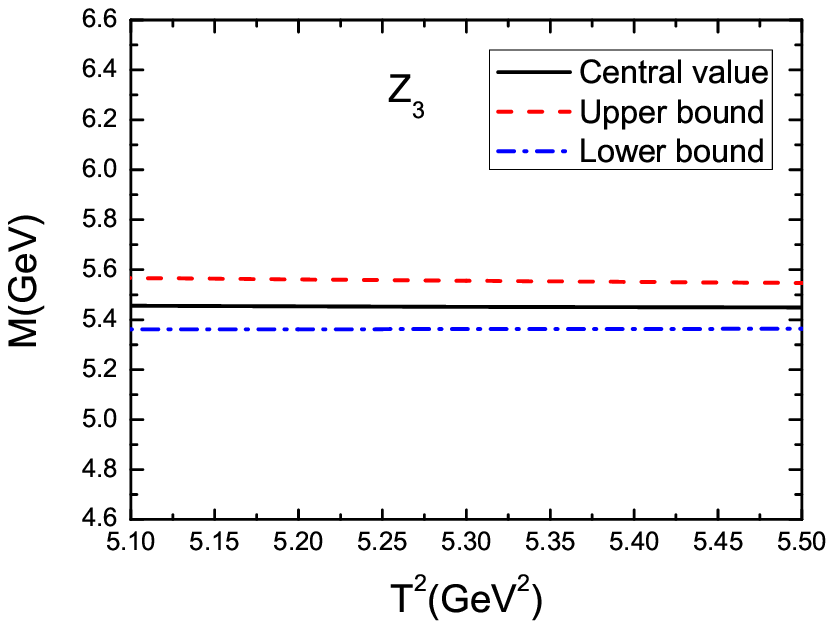}
\includegraphics[totalheight=5cm,width=7cm]{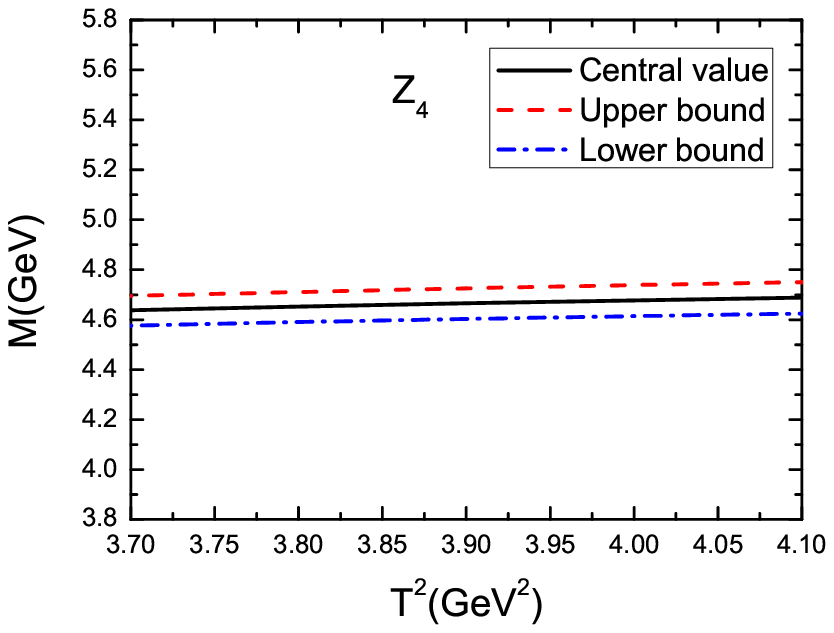}
\caption{The masses with variations of the Borel parameters $T^{2}$.}\label{fig:fig3}
\end{figure}

\begin{figure}[htp]
\centering
\includegraphics[totalheight=5cm,width=7cm]{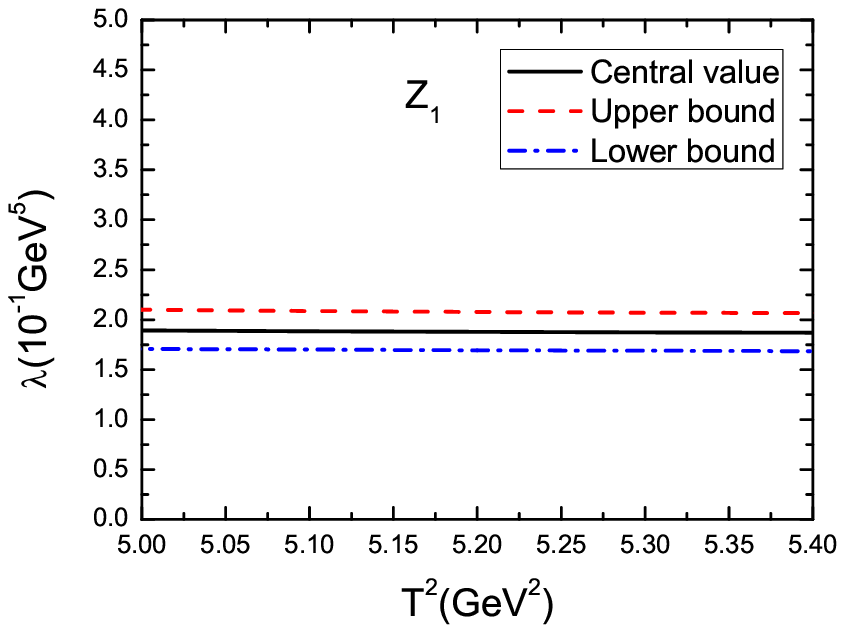}
\includegraphics[totalheight=5cm,width=7cm]{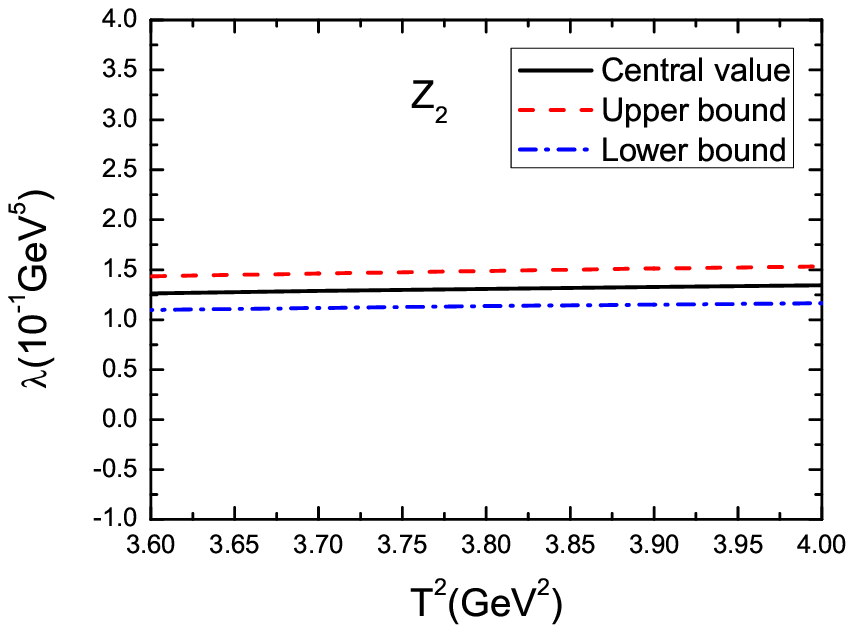}
\includegraphics[totalheight=5cm,width=7cm]{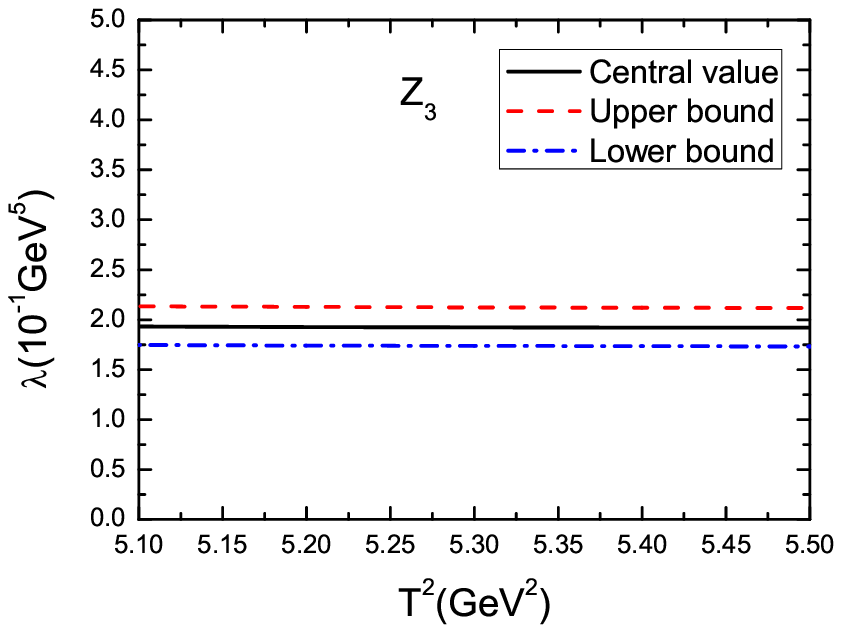}
\includegraphics[totalheight=5cm,width=7cm]{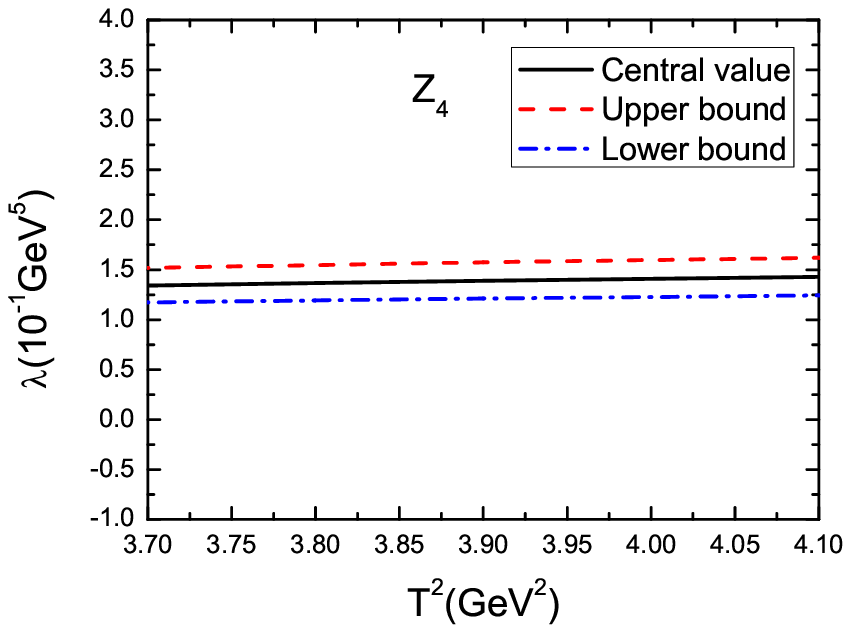}
\caption{The pole residues with variations of the Borel parameters $T^{2}$.}\label{fig:fig4}
\end{figure}

In Fig.\ref{fig:fig2}, the contributions of different condensate
terms in the operator product expansion are plotted with  the Borel parameters $T^{2}$ for the continuum
thresholds  $\sqrt{s_{0}}=5.90\,\text{GeV}$,
$5.20\,\text{GeV}$, $5.95\,\text{GeV}$ and
$5.25\,\text{GeV}$ at the  energy scales
$\mu=4.00\,\text{GeV}$, $2.90\,\text{GeV}$, $4.05\,\text{GeV}$
and $2.95\,\text{GeV}$ for the tetraquark states $Z_{1}$, $Z_{2}$, $Z_{3}$, $Z_{4}$, respectively. From the figure, we can
see the  contributions $D_i$ explicitly   and   choose reasonable Borel parameters. We take the
  $Z_{1}$ state as an example to illustrate the procedure. In that case, we observe that the perturbative term
and the $\langle\bar{q}q\rangle$ term play an important role, while the
other condensate terms play a minor important role. At the value
$T^{2}\leq4.9\,\text{GeV}^{2}$, the perturbative term and
$\langle\bar{q}q\rangle$ term  decrease monotonously and quickly with
increase of the $T^{2}$, which cannot lead to stable masses and pole residues. At the
value $T^{2}=(5.0-5.4)\,\text{GeV}^{2}$, the convergent behavior in
the operator product expansion is very good and the perturbative term makes the main contribution.
We present the optimal energy scales $\mu$,  ideal Borel parameters $T^2$, continuum threshold parameters $s_0$ and  pole contributions in Table 1. From the Table, we can see that the two  criteria of the QCD sum rules can be satisfied.

We take into account all uncertainties  of the input    parameters,
and obtain  the masses and pole residues of
 the  hidden-charm tetraquark states, which are shown explicitly in Figs.\ref{fig:fig3}-\ref{fig:fig4} and
Table \ref{tab:tablenotes}. From  Figs.\ref{fig:fig3}-\ref{fig:fig4},  we  can see that the Borel platforms exist. On the other hand, from Table \ref{tab:tablenotes}, we can see that the energy scale formula $ \mu =\sqrt{M_{X/Y/Z}^2-({2\mathbb{M}}_c)^2}$ is well satisfied. We expect to make reasonable predictions, the present predictions can be confronted with the experimental data in the future.

\begin{table}[!ht]
\begin{center}
\begin{tabular}{|c|c|c|c|c|c|c|}
  \hline
   \hline
  $J^{PC}$& $\mu(\text{GeV})$ & $T^{2}(\text{GeV}^{2})$ & $\sqrt{s_{0}}(\text{GeV})$  & pole & $M_{Z}(\text{GeV})$ &  $\lambda_{Z}(\text{GeV}^{5})$\\
   \hline
   $0^{++}(Z_{1})$ &$4.00$& $5.0-5.4$ & $5.90\pm0.10$ & $\left(42-57\right)\%$ & $5.43_{-0.09}^{+0.13}$ & $\left(1.88_{-0.19}^{+0.22}\right)\times10^{-1}$\\
   \hline
   $0^{++}(Z_{2})$ &$2.90$& $3.6-4.0$ & $5.20\pm0.10$ & $\left(42-61\right)\%$ & $4.64_{-0.09}^{+0.09}$ & $\left(1.31_{-0.21}^{+0.22}\right)\times10^{-1}$\\
   \hline
   $0^{++}(Z_{3})$ &$4.05$& $5.1-5.5$ & $5.95\pm0.10$ & $\left(43-57\right)\%$ & $5.45_{-0.09}^{+0.12}$ & $\left(1.92_{-0.19}^{+0.21}\right)\times10^{-1}$\\
   \hline
   $0^{++}(Z_{4})$ &$2.95$& $3.7-4.1$ & $5.25\pm0.10$ & $\left(43-61\right)\%$ & $4.67_{-0.09}^{+0.08}$ & $\left(1.39_{-0.22}^{+0.23}\right)\times10^{-1}$\\
  \hline
  \hline
\end{tabular}
\end{center}
\caption{The energy scales, Borel parameters, continuum threshold
parameters, pole contributions, masses and pole
residues for the scalar tetraquark states.}\label{tab:tablenotes}
\end{table}

In the following, we discuss the possible hadronic decay patterns of the $cu\bar{c}\bar{d}$ and
$cu\bar{c}\bar{s}$ scalar tetraquark states. Being composed of a diquark and antidiquark pair, a hidden-charm tetraquark state can decay very easily into a pair of open-charm $D$ mesons or one charmonium state plus a light meson through quark rearrangement. The strong decays are Okubo-Zweig-Iizuka super-allowed.
Under the restrictions of the symmetries and the masses of the studied scalar tetraquark states obtained in Table \ref{tab:tablenotes},
the possible two-body strong decay channels are
\begin{eqnarray}
Z_1&\longrightarrow& \chi_{c0}a_0^{+}(980),\, \eta_c\pi^+,\, J/\psi\rho^+(770),\, \psi(3770)\rho^+(770),\, \chi_{c1}a_1^+(1260),\,\nonumber\\
    &&\bar{D}_0^{*0}(2400)D_0^{*+}(2400),\, \bar{D}^0D^+,\, \bar{D}^{*0}(2007)D^{*+}(2010),\, \bar{D}_1^0(2420)D_1^+(2420),\nonumber\\
    &&\bar{D}_1^0(2430)D_1^+(2430),\nonumber\\
Z_2&\longrightarrow& \chi_{c0}a_0^{+}(980),\, \eta_c\pi^+,\, J/\psi\rho^+(770),\, \bar{D}^0D^+,\, \bar{D}^{*0}(2007)D^{*+}(2010),\,\nonumber\\
Z_3&\longrightarrow& \chi_{c0}K_0^{*+}(800),\, \eta_cK^+,\, J/\psi K^{*+}(892),\, \psi(3770)K^{*+}(892),\nonumber\\
&&\bar{D}_0^{*0}(2400)D_{s0}^{*+}(2317),\, \bar{D}^0D_s^+,\, \bar{D}^{*0}(2007)D_s^{*+},\, \bar{D}_1^0(2420)D_{s1}^+(2536),\nonumber\\ &&\bar{D}_1^0(2430)D_{s1}^+(2460),\nonumber\\
Z_4&\longrightarrow& \chi_{c0}K_0^{*+}(800),\, \eta_cK^+,\, J/\psi K^{*+}(892),\, \bar{D}^0D_s^+,\, \bar{D}^{*0}(2007)D_s^{*+},
\end{eqnarray}
which are kinematically allowed.
Theoretically, the mass of the ground state is lighter than the counterpart of its excited state, and the corresponding phase space is larger, thus the decay width of the resonance to the ground state is wider. This means that the ground state decay modes of the resonance can take place more easily. Besides, the excited sate is unstable, which bring some difficulty to the observation of the decay process for the resonance state.
Consequently, the dominant decay modes of the scalar tetraquark states are $Z_1(Z_2)\longrightarrow \eta_c\pi^+,\, J/\psi\rho^+(770),\, \bar{D}^0D^+$, $Z_3(Z_4)\longrightarrow \eta_cK^+,\, J/\psi K^{*+}(892),\, \bar{D}^0D_s^+.\,$
We can search for those scalar hidden-charm tetraquark states in those decay channels.

\section{Conclusion}
In this article, we study the
pseudoscalar-diquark-pseudoscalar-antidiquark type and
vector-diquark-vector-antidiquark type scalar hidden-charm
$cu\bar{c}\bar{d}$ ($cu\bar{c}\bar{s}$) tetraquark states with the
QCD sum rules by calculating the contributions of the vacuum
condensates up to dimension-10 in the operator product expansion. In
numerical calculations, we use the energy scale formula
$\mu=\sqrt{M_{X/Y/Z}^{2}-(2{\mathbb{M}}_{c})^{2}}$ to determine the
ideal energy scales of the QCD spectral densities and search for the
optimal Borel parameters $T^{2}$ and continuum thresholds  $s_{0}$
to satisfy the two criteria of the QCD sum rules (i.e. pole
dominance on the phenomenological side and convergence of the
operator product expansion). We obtain the masses and pole residues
of the scalar hidden-charm $cu\bar{c}\bar{d}$ ($cu\bar{c}\bar{s}$)
tetraquark states. The predicted masses are around
$5.43-5.45\,\text{GeV}$ for the $C\otimes C$ type tetraquark states
and $4.64-4.67\,\text{GeV}$ for the $C\gamma_{\mu}\gamma_{5}\otimes
\gamma_{5}\gamma^{\mu}C$ type ones, which can be confronted with the
experimental data in the future. Moreover, we discuss the possible
hadronic decay patterns of the two types of tetraquark states, and
list their dominant decays. As the predicted masses of the
$C\gamma_{\mu}\gamma_{5}\otimes \gamma_{5}\gamma^{\mu}C$ type
tetraquark states are lighter than the counterparts of the $C\otimes
C$ type ones and the two types of tetraquark states have the same
dominant decays, the widths of the $C\gamma_{\mu}\gamma_{5}\otimes
\gamma_{5}\gamma^{\mu}C$ type tetraquark states are narrower.
Therefore, the $C\gamma_{\mu}\gamma_{5}\otimes
\gamma_{5}\gamma^{\mu}C$ type tetraquark states can be observed more
easily, which can be testified in the future at the BESIII, LHCb and
Belle-II.

\section*{Acknowledgements}
This  work is supported by National Natural Science Foundation, Grant Number 11375063; the Fundamental Research Funds for the Central Universities, Grant Numbers 2016MS155 and 2016MS133.

\section*{Appendix}

The explicit expressions of the QCD spectral densities $\rho^{1,2,3,4}(s)$,
\begin{eqnarray}
\rho^{1}(s)&=&\rho^{3}(s)|_{m_{s}\rightarrow0,\langle\bar{s}s\rangle\rightarrow\langle\bar{q}q\rangle,\langle\bar{s}g_{s}\sigma Gs\rangle\rightarrow\langle\bar{q}g_{s}\sigma Gq\rangle}\ , \nonumber\\
\rho^{2}(s)&=&\rho^{4}(s)|_{m_{s}\rightarrow0,\langle\bar{s}s\rangle\rightarrow\langle\bar{q}q\rangle,\langle\bar{s}g_{s}\sigma Gs\rangle\rightarrow\langle\bar{q}g_{s}\sigma Gq\rangle}\ ,
\end{eqnarray}

\begin{eqnarray}
\rho_{0}^{3}(s)&=&\frac{1}{512\pi^{6}}\int_{y_{i}}^{y_{f}}dy\int_{z_{i}}^{1-y}dz\ yz(1-y-z)^{3}(s-\hat{m}_{c}^{2})^{2}(7s^{2}-6s\hat{m}_{c}^{2}+\hat{m}_{c}^{4}) \nonumber\\
&&-\frac{m_{s}m_{c}}{512\pi^{6}}\int_{y_{i}}^{y_{f}}dy\int_{z_{i}}^{1-y}dz\ (y+z)(1-y-z)^{2}(s-\hat{m}_{c}^{2})^{2}(5s-2\hat{m}_{c}^{2})\ ,
\end{eqnarray}

\begin{eqnarray}
\rho_{0}^{4}(s)&=&\frac{1}{256\pi^{6}}\int_{y_{i}}^{y_{f}}dy\int_{z_{i}}^{1-y}dz\ yz(1-y-z)^{3}(s-\hat{m}_{c}^{2})^{2}(7s^{2}-6s\hat{m}_{c}^{2}+\hat{m}_{c}^{4}) \nonumber\\
&&+\frac{1}{256\pi^{6}}\int_{y_{i}}^{y_{f}}dy\int_{z_{i}}^{1-y}dz\ yz(1-y-z)^{2}(s-\hat{m}_{c}^{2})^{3}(3s-\hat{m}_{c}^{2}) \nonumber\\
&&-\frac{m_{s}m_{c}}{256\pi^{6}}\int_{y_{i}}^{y_{f}}dy\int_{z_{i}}^{1-y}dz\ (y+z)(1-y-z)^{2}(s-\hat{m}_{c}^{2})^{2}(5s-2\hat{m}_{c}^{2})\ ,
\end{eqnarray}

\begin{eqnarray}
\rho_{3}^{3}(s)&=&\frac{m_{c}\left(\langle\bar{q}q\rangle+\langle\bar{s}s\rangle\right)}{32\pi^{4}}\int_{y_{i}}^{y_{f}}dy\int_{z_{i}}^{1-y}dz\ (y+z)(1-y-z)(s-\hat{m}_{c}^{2})(2s-\hat{m}_{c}^{2}) \nonumber\\
&&+\frac{m_{s}\langle\bar{s}s\rangle}{32\pi^{4}}\int_{y_{i}}^{y_{f}}dy\int_{z_{i}}^{1-y}dz\ yz(1-y-z)(10s^{2}-12s\hat{m}_{c}^{2}+3\hat{m}_{c}^{4}) \nonumber\\
&&-\frac{m_{s}m_{c}^{2}\langle\bar{q}q\rangle}{16\pi^{4}}\int_{y_{i}}^{y_{f}}dy\int_{z_{i}}^{1-y}dz\ (s-\hat{m}_{c}^{2})\ ,
\end{eqnarray}

\begin{eqnarray}
\rho_{3}^{4}(s)&=&\frac{m_{c}\left(\langle\bar{q}q\rangle+\langle\bar{s}s\rangle\right)}{16\pi^{4}}\int_{y_{i}}^{y_{f}}dy\int_{z_{i}}^{1-y}dz\ (y+z)(1-y-z)(s-\hat{m}_{c}^{2})(2s-\hat{m}_{c}^{2}) \nonumber\\
&&-\frac{m_{s}m_{c}^{2}\langle\bar{q}q\rangle}{4\pi^{4}}\int_{y_{i}}^{y_{f}}dy\int_{z_{i}}^{1-y}dz\ (s-\hat{m}_{c}^{2}) \nonumber\\
&&+\frac{m_{s}\langle\bar{s}s\rangle}{16\pi^{4}}\int_{y_{i}}^{y_{f}}dy\int_{z_{i}}^{1-y}dz\ yz(1-y-z)(10s^{2}-12s\hat{m}_{c}^{2}+3\hat{m}_{c}^{4}) \nonumber\\
&&+\frac{m_{s}\langle\bar{s}s\rangle}{16\pi^{4}}\int_{y_{i}}^{y_{f}}dy\int_{z_{i}}^{1-y}dz\ yz(s-\hat{m}_{c}^{2})(2s-\hat{m}_{c}^{2})\ ,
\end{eqnarray}

\begin{eqnarray}
\rho_{4}^{3}(s)&=&-\frac{m_{c}^{2}}{384\pi^{4}}\langle\frac{\alpha_{s}GG}{\pi}\rangle\int_{y_{i}}^{y_{f}}dy\int_{z_{i}}^{1-y}dz\ \left(\frac{y}{z^{2}}+\frac{z}{y^{2}}\right)(1-y-z)^{3} \left\{(2s-\hat{m}_{c}^{2})+\frac{\hat{m}_{c}^{4}}{6}\delta(s-\hat{m}_{c}^{2})\right\} \nonumber\\
&&+\frac{1}{512\pi^{4}}\langle\frac{\alpha_{s}GG}{\pi}\rangle\int_{y_{i}}^{y_{f}}dy\int_{z_{i}}^{1-y}dz\ (y+z)(1-y-z)^{2}(10s^{2}-12s\hat{m}_{c}^{2}+3\hat{m}_{c}^{4}) \nonumber\\
&&-\frac{m_{s}m_{c}}{512\pi^{4}}\langle\frac{\alpha_{s}GG}{\pi}\rangle\int_{y_{i}}^{y_{f}}dy\int_{z_{i}}^{1-y}dz\ \left(\frac{y}{z^{2}}+\frac{z}{y^{2}}\right)(1-y-z)^{2}(3s-2\hat{m}_{c}^{2}) \nonumber\\
&&+\frac{m_{s}m_{c}^{3}}{768\pi^{4}}\langle\frac{\alpha_{s}GG}{\pi}\rangle\int_{y_{i}}^{y_{f}}dy\int_{z_{i}}^{1-y}dz\ \left(\frac{y}{z^{3}}+\frac{z}{y^{3}}+\frac{1}{z^{2}}+\frac{1}{y^{2}}\right)(1-y-z)^{2} \left\{1+\frac{\hat{m}_{c}^{2}}{2}\delta(s-\hat{m}_{c}^{2})\right\} \nonumber\\
&&-\frac{m_{s}m_{c}}{256\pi^{4}}\langle\frac{\alpha_{s}GG}{\pi}\rangle\int_{y_{i}}^{y_{f}}dy\int_{z_{i}}^{1-y}dz\ (1-y-z)(3s-2\hat{m}_{c}^{2})\ ,
\end{eqnarray}

\begin{eqnarray}
\rho_{4}^{4}(s)&=&-\frac{m_{c}^{2}}{192\pi^{4}}\langle\frac{\alpha_{s}GG}{\pi}\rangle\int_{y_{i}}^{y_{f}}dy\int_{z_{i}}^{1-y}dz\ \left(\frac{y}{z^{2}}+\frac{z}{y^{2}}\right)(1-y-z)^{3}\left\{(2s-\hat{m}_{c}^{2})+\frac{\hat{m}_{c}^{4}}{6}\delta(s-\hat{m}_{c}^{2})\right\} \nonumber\\
&&-\frac{m_{c}^{2}}{384\pi^{4}}\langle\frac{\alpha_{s}GG}{\pi}\rangle\int_{y_{i}}^{y_{f}}dy\int_{z_{i}}^{1-y}dz\ \left(\frac{y}{z^{2}}+\frac{z}{y^{2}}\right)(1-y-z)^{2}(3s-2\hat{m}_{c}^{2}) \nonumber\\
&&+\frac{1}{768\pi^{4}}\langle\frac{\alpha_{s}GG}{\pi}\rangle\int_{y_{i}}^{y_{f}}dy\int_{z_{i}}^{1-y}dz\ (y+z)(1-y-z)^{2}(10s^{2}-12s\hat{m}_{c}^{2}+3\hat{m}_{c}^{4}) \nonumber\\
&&+\frac{1}{384\pi^{4}}\langle\frac{\alpha_{s}GG}{\pi}\rangle\int_{y_{i}}^{y_{f}}dy\int_{z_{i}}^{1-y}dz\ (y+z)(1-y-z)(s-\hat{m}_{c}^{2})(2s-\hat{m}_{c}^{2}) \nonumber\\
&&+\frac{1}{3456\pi^{4}}\langle\frac{\alpha_{s}GG}{\pi}\rangle\int_{y_{i}}^{y_{f}}dy\int_{z_{i}}^{1-y}dz\ (1-y-z)^{3}(10s^{2}-12s\hat{m}_{c}^{2}+3\hat{m}_{c}^{4}) \nonumber\\
&&+\frac{1}{576\pi^{4}}\langle\frac{\alpha_{s}GG}{\pi}\rangle\int_{y_{i}}^{y_{f}}dy\int_{z_{i}}^{1-y}dz\ yz(1-y-z)(10s^{2}-12s\hat{m}_{c}^{2}+3\hat{m}_{c}^{4}) \nonumber\\
&&+\frac{1}{576\pi^{4}}\langle\frac{\alpha_{s}GG}{\pi}\rangle\int_{y_{i}}^{y_{f}}dy\int_{z_{i}}^{1-y}dz\ (1-y-z)^{2}(s-\hat{m}_{c}^{2})(2s-\hat{m}_{c}^{2}) \nonumber\\
&&+\frac{1}{288\pi^{4}}\langle\frac{\alpha_{s}GG}{\pi}\rangle\int_{y_{i}}^{y_{f}}dy\int_{z_{i}}^{1-y}dz\ yz(s-\hat{m}_{c}^{2})(2s-\hat{m}_{c}^{2}) \nonumber\\
&&-\frac{m_{s}m_{c}}{256\pi^{4}}\langle\frac{\alpha_{s}GG}{\pi}\rangle\int_{y_{i}}^{y_{f}}dy\int_{z_{i}}^{1-y}dz\ \left(\frac{z}{y^{2}}+\frac{y}{z^{2}}\right)(1-y-z)^{2}(3s-2\hat{m}_{c}^{2}) \nonumber\\
&&+\frac{m_{s}m_{c}^{3}}{384\pi^{4}}\langle\frac{\alpha_{s}GG}{\pi}\rangle\int_{y_{i}}^{y_{f}}dy\int_{z_{i}}^{1-y}dz\ \left(\frac{1}{y^{2}}+\frac{1}{z^{2}}+\frac{y}{z^{3}}+\frac{z}{y^{3}}\right)(1-y-z)^{2}\left\{1+\frac{\hat{m}_{c}^{2}}{2}\delta(s-\hat{m}_{c}^{2})\right\} \nonumber\\
&&+\frac{m_{s}m_{c}}{128\pi^{4}}\langle\frac{\alpha_{s}GG}{\pi}\rangle\int_{y_{i}}^{y_{f}}dy\int_{z_{i}}^{1-y}dz\ (1-y-z)(3s-2\hat{m}_{c}^{2}) \nonumber\\
&&-\frac{m_{s}m_{c}}{384\pi^{4}}\langle\frac{\alpha_{s}GG}{\pi}\rangle\int_{y_{i}}^{y_{f}}dy\int_{z_{i}}^{1-y}dz\ \left(\frac{y}{z}+\frac{z}{y}\right)(1-y-z)(3s-2\hat{m}_{c}^{2}) \nonumber\\
&&-\frac{m_{s}m_{c}}{768\pi^{4}}\langle\frac{\alpha_{s}GG}{\pi}\rangle\int_{y_{i}}^{y_{f}}dy\int_{z_{i}}^{1-y}dz\ \left(\frac{1}{y}+\frac{1}{z}\right)(1-y-z)^{2}(3s-2\hat{m}_{c}^{2})\ ,
\end{eqnarray}

\begin{eqnarray}
\rho_{5}^{3}(s)&=&-\frac{m_{c}\left(\langle\bar{q}g_{s}\sigma Gq\rangle+\langle\bar{s}g_{s}\sigma Gs\rangle\right)}{128\pi^{4}}\int_{y_{i}}^{y_{f}}dy\int_{z_{i}}^{1-y}dz\ (y+z)(3s-2\hat{m}_{c}^{2}) \nonumber\\
&&+\frac{m_{c}\left(\langle\bar{q}g_{s}\sigma Gq\rangle+\langle\bar{s}g_{s}\sigma Gs\rangle\right)}{128\pi^{4}}\int_{y_{i}}^{y_{f}}dy\int_{z_{i}}^{1-y}dz\ \left(\frac{y}{z}+\frac{z}{y}\right)(1-y-z)(3s-2\hat{m}_{c}^{2}) \nonumber\\
&&-\frac{m_{s}\langle\bar{s}g_{s}\sigma Gs\rangle}{32\pi^{4}}\int_{y_{i}}^{y_{f}}dy\int_{z_{i}}^{1-y}dz\ yz\left\{(2s-\hat{m}_{c}^{2})+\frac{\hat{m}_{c}^{4}}{6}\delta(s-\hat{m}_{c}^{2})\right\} \nonumber\\
&&+\frac{m_{s}m_{c}^{2}\langle\bar{q}g_{s}\sigma Gq\rangle}{64\pi^{4}}\int_{y_{i}}^{y_{f}}dy\ -\frac{m_{s}m_{c}^{2}\langle\bar{q}g_{s}\sigma Gq\rangle}{128\pi^{4}}\int_{y_{i}}^{y_{f}}dy\int_{z_{i}}^{1-y}dz\ \left(\frac{1}{y}+\frac{1}{z}\right)\ ,
\end{eqnarray}

\begin{eqnarray}
\rho_{5}^{4}(s)&=&-\frac{m_{c}\left(\langle\bar{q}g_{s}\sigma Gq\rangle+\langle\bar{s}g_{s}\sigma Gs\rangle\right)}{96\pi^{4}}\int_{y_{i}}^{y_{f}}dy\int_{z_{i}}^{1-y}dz\ (y+z)(3s-2\hat{m}_{c}^{2}) \nonumber\\
&&+\frac{m_{c}\left(\langle\bar{q}g_{s}\sigma Gq\rangle+\langle\bar{s}g_{s}\sigma Gs\rangle\right)}{96\pi^{4}}\int_{y_{i}}^{y_{f}}dy\int_{z_{i}}^{1-y}dz\ (1-y-z)(3s-2\hat{m}_{c}^{2}) \nonumber\\
&&-\frac{m_{s}\langle\bar{s}g_{s}\sigma Gs\rangle}{16\pi^{4}}\int_{y_{i}}^{y_{f}}dy\int_{z_{i}}^{1-y}dz\ yz\left\{(2s-\hat{m}_{c}^{2})+\frac{\hat{m}_{c}^{4}}{6}\delta(s-\hat{m}_{c}^{2})\right\} \nonumber\\
&&+\frac{m_{s}m_{c}^{2}\langle\bar{q}g_{s}\sigma Gq\rangle}{16\pi^{4}}\int_{y_{i}}^{y_{f}}dy -\frac{m_{s}\langle\bar{s}g_{s}\sigma Gs\rangle}{96\pi^{4}}\int_{y_{i}}^{y_{f}}dy\ y(1-y)(3s-2\tilde{m}_{c}^{2}) \nonumber\\
&&-\frac{m_{s}m_{c}^{2}\langle\bar{q}g_{s}\sigma Gq\rangle}{96\pi^{4}}\int_{y_{i}}^{y_{f}}dy\int_{z_{i}}^{1-y}dz\ \left(\frac{1}{y}+\frac{1}{z}\right)\ ,
\end{eqnarray}

\begin{eqnarray}
\rho_{6}^{3}(s)&=& \frac{m_{c}^{2}\langle\bar{q}q\rangle\langle\bar{s}s\rangle}{12\pi^{2}}\int_{y_{i}}^{y_{f}}dy+\frac{g_{s}^{2}\left(\langle\bar{q}q\rangle^{2} +\langle\bar{s}s\rangle^{2}\right)}{216\pi^{4}}\int_{y_{i}}^{y_{f}}dy\int_{z_{i}}^{1-y}dz\ yz\left\{(2s-\hat{m}_{c}^{2})+\frac{\hat{m}_{c}^{4}}{6}\delta(s-\hat{m}_{c}^{2})\right\} \nonumber\\
&&-\frac{7g_{s}^{2}\left(\langle\bar{q}q\rangle^{2}+\langle\bar{s}s\rangle^{2}\right)}{2592\pi^{4}}\int_{y_{i}}^{y_{f}}dy\int_{z_{i}}^{1-y}dz\ \left(\frac{y}{z}+\frac{z}{y}\right)(1-y-z)(3s-2\hat{m}_{c}^{2}) \nonumber\\
&&-\frac{5m_{c}^{2}g_{s}^{2}\left(\langle\bar{q}q\rangle^{2}+\langle\bar{s}s\rangle^{2}\right)}{1944\pi^{4}}\int_{y_{i}}^{y_{f}}dy\int_{z_{i}}^{1-y}dz\ \left(\frac{y}{z^{2}}+\frac{z}{y^{2}}\right)(1-y-z)\left\{1+\frac{\hat{m}_{c}^{2}}{2}\delta(s-\hat{m}_{c}^{2})\right\} \nonumber\\
&&-\frac{g_{s}^{2}\left(\langle\bar{q}q\rangle^{2}+\langle\bar{s}s\rangle^{2}\right)}{648\pi^{4}}\int_{y_{i}}^{y_{f}}dy\int_{z_{i}}^{1-y}dz\ (y+z)(1-y-z)\left\{(2s-\hat{m}_{c}^{2})+\frac{\hat{m}_{c}^{4}}{6}\delta(s-\hat{m}_{c}^{2})\right\} \nonumber\\
&&+\frac{m_{s}m_{c}\langle\bar{q}q\rangle\langle\bar{s}s\rangle}{48\pi^{2}}\int_{y_{i}}^{y_{f}}dy\ \bigg\{2+\tilde{m}_{c}^{2}\delta(s-\tilde{m}_{c}^{2})\bigg\} -\frac{m_{s}m_{c}g_{s}^{2}\langle\bar{q}q\rangle^{2}}{2592\pi^{4}}\int_{y_{i}}^{y_{f}}dy\ \bigg\{2+\tilde{m}_{c}^{2}\delta(s-\tilde{m}_{c}^{2})\bigg\} \nonumber\\
&&+\frac{m_{s}m_{c}^{3}g_{s}^{2}\langle\bar{q}q\rangle^{2}}{864\pi^{4}}\int_{y_{i}}^{y_{f}}dy\int_{z_{i}}^{1-y}dz\ \left(\frac{1}{y^{2}}+\frac{1}{z^{2}}\right)\delta(s-\hat{m}_{c}^{2}) \nonumber\\ &&+\frac{m_{s}m_{c}g_{s}^{2}\langle\bar{q}q\rangle^{2}}{432\pi^{4}}\int_{y_{i}}^{y_{f}}dy\int_{z_{i}}^{1-y}dz\ \left(\frac{1}{y}+\frac{1}{z}\right) \nonumber\\
&&+\frac{m_{s}m_{c}g_{s}^{2}\langle\bar{q}q\rangle^{2}}{2592\pi^{4}}\int_{y_{i}}^{y_{f}}dy\int_{z_{i}}^{1-y}dz\ \left(\frac{z}{y}+\frac{y}{z}\right)\bigg\{2+\hat{m}_{c}^{2}\delta(s-\hat{m}_{c}^{2})\bigg\}\ ,
\end{eqnarray}

\begin{eqnarray}
\rho_{6}^{4}(s)&=& \frac{m_{c}^{2}\langle\bar{q}q\rangle\langle\bar{s}s\rangle}{3\pi^{2}}\int_{y_{i}}^{y_{f}}dy+\frac{g_{s}^{2} \left(\langle\bar{q}q\rangle^{2}+\langle\bar{s}s\rangle^{2}\right)}{108\pi^{4}}\int_{y_{i}}^{y_{f}}dy\int_{z_{i}}^{1-y}dz\ yz\left\{(2s-\hat{m}_{c}^{2})+\frac{\hat{m}_{c}^{4}}{6}\delta(s-\hat{m}_{c}^{2})\right\} \nonumber\\
&&+\frac{g_{s}^{2}\left(\langle\bar{q}q\rangle^{2}+\langle\bar{s}s\rangle^{2}\right)}{648\pi^{4}}\int_{y_{i}}^{y_{f}}dy\ y(1-y)(3s-2\tilde{m}_{c}^{2}) \nonumber\\
&&-\frac{5m_{c}^{2}g_{s}^{2}\left(\langle\bar{q}q\rangle^{2}+\langle\bar{s}s\rangle^{2}\right)}{972\pi^{4}}\int_{y_{i}}^{y_{f}}dy\int_{z_{i}}^{1-y}dz\ \left(\frac{y}{z^{2}}+\frac{z}{y^{2}}\right)(1-y-z)\left\{1+\frac{\hat{m}_{c}^{2}}{2}\delta(s-\hat{m}_{c}^{2})\right\} \nonumber\\
&&-\frac{g_{s}^{2}\left(\langle\bar{q}q\rangle^{2}+\langle\bar{s}s\rangle^{2}\right)}{162\pi^{4}}\int_{y_{i}}^{y_{f}}dy\int_{z_{i}}^{1-y}dz\ \left(\frac{y}{z}+\frac{z}{y}\right)(1-y-z)(3s-2\hat{m}_{c}^{2}) \nonumber\\
&&-\frac{g_{s}^{2}\left(\langle\bar{q}q\rangle^{2}+\langle\bar{s}s\rangle^{2}\right)}{81\pi^{4}}\int_{y_{i}}^{y_{f}}dy\int_{z_{i}}^{1-y}dz\ (y+z)(1-y-z)\left\{(2s-\hat{m}_{c}^{2})+\frac{\hat{m}_{c}^{4}}{6}\delta(s-\hat{m}_{c}^{2})\right\} \nonumber\\
&&+\frac{m_{s}m_{c}\langle\bar{q}q\rangle\langle\bar{s}s\rangle}{24\pi^{2}}\int_{y_{i}}^{y_{f}}dy\ \bigg\{2+\tilde{m}_{c}^{2}\delta(s-\tilde{m}_{c}^{2})\bigg\} -\frac{m_{s}m_{c}g_{s}^{2}\langle\bar{q}q\rangle^{2}}{1296\pi^{4}}\int_{y_{i}}^{y_{f}}dy\ \bigg\{2+\tilde{m}_{c}^{2}\delta(s-\tilde{m}_{c}^{2})\bigg\} \nonumber\\
&&+\frac{m_{s}m_{c}^{3}g_{s}^{2}\langle\bar{q}q\rangle^{2}}{432\pi^{4}}\int_{y_{i}}^{y_{f}}dy\int_{z_{i}}^{1-y}dz\ \left(\frac{1}{y^{2}}+\frac{1}{z^{2}}\right)\delta(s-\hat{m}_{c}^{2}) \nonumber\\
&&+\frac{m_{s}m_{c}g_{s}^{2}\langle\bar{q}q\rangle^{2}}{216\pi^{4}}\int_{y_{i}}^{y_{f}}dy\int_{z_{i}}^{1-y}dz\ \left(\frac{1}{y}+\frac{1}{z}\right) \nonumber\\
&&+\frac{m_{s}m_{c}g_{s}^{2}\langle\bar{q}q\rangle^{2}}{1296\pi^{4}}\int_{y_{i}}^{y_{f}}dy\int_{z_{i}}^{1-y}dz\ \left(\frac{z}{y}+\frac{y}{z}\right)\bigg\{2+\hat{m}_{c}^{2}\delta(s-\hat{m}_{c}^{2})\bigg\}\ ,
\end{eqnarray}

\begin{eqnarray}
\rho_{7}^{3}(s)&=& -\frac{m_{c}^{3}\left(\langle\bar{q}q\rangle+\langle\bar{s}s\rangle\right)}{576\pi^{2}}\langle\frac{\alpha_{s}GG}{\pi}\rangle\int_{y_{i}}^{y_{f}}dy\int_{z_{i}}^{1-y}dz\ \left(\frac{1}{y^{2}}+\frac{1}{z^{2}}+\frac{y}{z^{3}}+\frac{z}{y^{3}}\right)(1-y-z) \nonumber\\
&&\left(1+\frac{\hat{m}_{c}^{2}}{T^{2}}\right)\delta(s-\hat{m}_{c}^{2}) \nonumber\\
&&+\frac{m_{c}\left(\langle\bar{q}q\rangle+\langle\bar{s}s\rangle\right)}{192\pi^{2}}\langle\frac{\alpha_{s}GG}{\pi}\rangle\int_{y_{i}}^{y_{f}}dy\int_{z_{i}}^{1-y}dz\ \left(\frac{y}{z^{2}}+\frac{z}{y^{2}}\right)(1-y-z)\bigg\{2+\hat{m}_{c}^{2}\delta(s-\hat{m}_{c}^{2})\bigg\} \nonumber\\
&&+\frac{m_{c}\left(\langle\bar{q}q\rangle+\langle\bar{s}s\rangle\right)}{192\pi^{2}}\langle\frac{\alpha_{s}GG}{\pi}\rangle\int_{y_{i}}^{y_{f}}dy\int_{z_{i}}^{1-y}dz\ \bigg\{2+\hat{m}_{c}^{2}\delta(s-\hat{m}_{c}^{2})\bigg\} \nonumber\\
&&+\frac{m_{c}\left(\langle\bar{q}q\rangle+\langle\bar{s}s\rangle\right)}{1152\pi^{2}}\langle\frac{\alpha_{s}GG}{\pi}\rangle\int_{y_{i}}^{y_{f}}dy\ \bigg\{2+\tilde{m}_{c}^{2}\delta(s-\tilde{m}_{c}^{2})\bigg\} \nonumber\\
&&-\frac{m_{s}m_{c}^{2}\langle\bar{s}s\rangle}{288\pi^{2}}\langle\frac{\alpha_{s}GG}{\pi}\rangle\int_{y_{i}}^{y_{f}}dy\int_{z_{i}}^{1-y}dz\ \left(\frac{y}{z^{2}}+\frac{z}{y^{2}}\right)(1-y-z)\left(1+\frac{\hat{m}_{c}^{2}}{T^{2}}+\frac{\hat{m}_{c}^{4}}{2T^{4}}\right)\delta(s-\hat{m}_{c}^{2}) \nonumber\\
&&-\frac{m_{s}m_{c}^{2}\langle\bar{q}q\rangle}{96\pi^{2}}\langle\frac{\alpha_{s}GG}{\pi}\rangle\int_{y_{i}}^{y_{f}}dy\int_{z_{i}}^{1-y}dz\ \left(\frac{1}{y^{2}}+\frac{1}{z^{2}}\right)\delta(s-\hat{m}_{c}^{2}) \nonumber\\
&&+\frac{m_{s}m_{c}^{4}\langle\bar{q}q\rangle}{288\pi^{2}T^{2}}\langle\frac{\alpha_{s}GG}{\pi}\rangle\int_{y_{i}}^{y_{f}}dy\int_{z_{i}}^{1-y}dz\ \left(\frac{1}{y^{3}}+\frac{1}{z^{3}}\right)\delta(s-\hat{m}_{c}^{2}) \nonumber\\
&&+\frac{m_{s}\langle\bar{s}s\rangle}{768\pi^{2}}\langle\frac{\alpha_{s}GG}{\pi}\rangle\int_{y_{i}}^{y_{f}}dy\int_{z_{i}}^{1-y}dz\ (y+z)\left\{6+4\hat{m}_{c}^{2}\delta(s-\hat{m}_{c}^{2})+\frac{\hat{m}_{c}^{4}\delta(s-\hat{m}_{c}^{2})}{T^{2}}\right\} \nonumber\\
&&-\frac{m_{s}m_{c}^{2}\langle\bar{q}q\rangle}{576\pi^{2}}\langle\frac{\alpha_{s}GG}{\pi}\rangle\int_{y_{i}}^{y_{f}}dy\ \left(1+\frac{\tilde{m}_{c}^{2}}{T^{2}}\right)\delta(s-\tilde{m}_{c}^{2})\ ,
\end{eqnarray}

\begin{eqnarray}
\rho_{7}^{4}(s)&=& -\frac{m_{c}^{3}(\langle\bar{q}q\rangle+\langle\bar{s}s\rangle)}{288\pi^{2}}\langle\frac{\alpha_{s}GG}{\pi}\rangle\int_{y_{i}}^{y_{f}}dy\int_{z_{i}}^{1-y}dz\ \left(\frac{1}{y^{2}}+\frac{1}{z^{2}}+\frac{y}{z^{3}}+\frac{z}{y^{3}}\right)(1-y-z)\nonumber\\
&&\left(1+\frac{\hat{m}_{c}^{2}}{T^{2}}\right)\delta(s-\hat{m}_{c}^{2}) \nonumber\\
&&+\frac{m_{c}(\langle\bar{q}q\rangle+\langle\bar{s}s\rangle)}{96\pi^{2}}\langle\frac{\alpha_{s}GG}{\pi}\rangle\int_{y_{i}}^{y_{f}}dy\int_{z_{i}}^{1-y}dz\ \left(\frac{y}{z^{2}}+\frac{z}{y^{2}}\right)(1-y-z)\bigg\{2+\hat{m}_{c}^{2}\delta(s-\hat{m}_{c}^{2})\bigg\} \nonumber\\
&&-\frac{m_{c}(\langle\bar{q}q\rangle+\langle\bar{s}s\rangle)}{96\pi^{2}}\langle\frac{\alpha_{s}GG}{\pi}\rangle\int_{y_{i}}^{y_{f}}dy\int_{z_{i}}^{1-y}dz\ \bigg\{2+\hat{m}_{c}^{2}\delta(s-\hat{m}_{c}^{2})\bigg\} \nonumber\\
&&+\frac{m_{c}(\langle\bar{q}q\rangle+\langle\bar{s}s\rangle)}{288\pi^{2}}\langle\frac{\alpha_{s}GG}{\pi}\rangle\int_{y_{i}}^{y_{f}}dy\int_{z_{i}}^{1-y}dz\ \left(\frac{y}{z}+\frac{z}{y}\right)\bigg\{2+\hat{m}_{c}^{2}\delta(s-\hat{m}_{c}^{2})\bigg\} \nonumber\\
&&+\frac{m_{c}(\langle\bar{q}q\rangle+\langle\bar{s}s\rangle)}{288\pi^{2}}\langle\frac{\alpha_{s}GG}{\pi}\rangle\int_{y_{i}}^{y_{f}}dy\int_{z_{i}}^{1-y}dz\ \left(\frac{1}{y}+\frac{1}{z}\right)(1-y-z)\bigg\{2+\hat{m}_{c}^{2}\delta(s-\hat{m}_{c}^{2})\bigg\} \nonumber\\
&&+\frac{m_{c}(\langle\bar{q}q\rangle+\langle\bar{s}s\rangle)}{576\pi^{2}}\langle\frac{\alpha_{s}GG}{\pi}\rangle\int_{y_{i}}^{y_{f}}dy\ \bigg\{2+\tilde{m}_{c}^{2}\delta(s-\tilde{m}_{c}^{2})\bigg\} \nonumber\\
&&-\frac{m_{s}m_{c}^{2}\langle\bar{q}q\rangle}{24\pi^{2}}\langle\frac{\alpha_{s}GG}{\pi}\rangle\int_{y_{i}}^{y_{f}}dy\int_{z_{i}}^{1-y}dz\ \left(\frac{1}{y^{2}}+\frac{1}{z^{2}}\right)\delta(s-\hat{m}_{c}^{2}) \nonumber\\
&&+\frac{m_{s}m_{c}^{4}\langle\bar{q}q\rangle}{72\pi^{2}T^{2}}\langle\frac{\alpha_{s}GG}{\pi}\rangle\int_{y_{i}}^{y_{f}}dy\int_{z_{i}}^{1-y}dz\ \left(\frac{1}{y^{3}}+\frac{1}{z^{3}}\right)\delta(s-\hat{m}_{c}^{2}) \nonumber\\
&&-\frac{m_{s}m_{c}^{2}\langle\bar{s}s\rangle}{144\pi^{2}}\langle\frac{\alpha_{s}GG}{\pi}\rangle\int_{y_{i}}^{y_{f}}dy\int_{z_{i}}^{1-y}dz\ \left(\frac{y}{z^{2}}+\frac{z}{y^{2}}\right)(1-y-z)\left(1+\frac{\hat{m}_{c}^{2}}{T^{2}}+\frac{\hat{m}_{c}^{4}}{2T^{4}}\right)\delta(s-\hat{m}_{c}^{2}) \nonumber\\
&&-\frac{m_{s}m_{c}^{2}\langle\bar{s}s\rangle}{288\pi^{2}}\langle\frac{\alpha_{s}GG}{\pi}\rangle\int_{y_{i}}^{y_{f}}dy\int_{z_{i}}^{1-y}dz\ \left(\frac{y}{z^{2}}+\frac{z}{y^{2}}\right)\left(1+\frac{\hat{m}_{c}^{2}}{T^{2}}\right)\delta(s-\hat{m}_{c}^{2}) \nonumber\\
&&+\frac{m_{s}\langle\bar{s}s\rangle}{192\pi^{2}}\langle\frac{\alpha_{s}GG}{\pi}\rangle\int_{y_{i}}^{y_{f}}dy\int_{z_{i}}^{1-y}dz\ (y+z) \nonumber\\
&&+\frac{m_{s}\langle\bar{s}s\rangle}{144\pi^{2}}\langle\frac{\alpha_{s}GG}{\pi}\rangle\int_{y_{i}}^{y_{f}}dy\int_{z_{i}}^{1-y}dz\ (1-y-z) \nonumber\\
&&+\frac{m_{s}\langle\bar{s}s\rangle}{288\pi^{2}}\langle\frac{\alpha_{s}GG}{\pi}\rangle\int_{y_{i}}^{y_{f}}dy\int_{z_{i}}^{1-y}dz\ (y+z)\left(\hat{m}_{c}^{2}+\frac{\hat{m}_{c}^{4}}{4T^{2}}\right)\delta(s-\hat{m}_{c}^{2}) \nonumber\\
&&+\frac{m_{s}\langle\bar{s}s\rangle}{216\pi^{2}}\langle\frac{\alpha_{s}GG}{\pi}\rangle\int_{y_{i}}^{y_{f}}dy\int_{z_{i}}^{1-y}dz\ (1-y-z)\left(\hat{m}_{c}^{2}+\frac{\hat{m}_{c}^{4}}{4T^{2}}\right)\delta(s-\hat{m}_{c}^{2}) \nonumber\\
&&+\frac{11m_{s}\langle\bar{s}s\rangle}{3456\pi^{2}}\langle\frac{\alpha_{s}GG}{\pi}\rangle\int_{y_{i}}^{y_{f}}dy\ \bigg\{2+\tilde{m}_{c}^{2}\delta(s-\tilde{m}_{c}^{2})\bigg\} \nonumber\\
&&-\frac{m_{s}m_{c}^{2}\langle\bar{q}q\rangle}{72\pi^{2}}\langle\frac{\alpha_{s}GG}{\pi}\rangle\int_{y_{i}}^{y_{f}}dy\int_{z_{i}}^{1-y}dz\ \frac{1}{yz}\delta(s-\hat{m}_{c}^{2}) \nonumber\\
&&-\frac{m_{s}m_{c}^{2}\langle\bar{q}q\rangle}{144\pi^{2}}\langle\frac{\alpha_{s}GG}{\pi}\rangle\int_{y_{i}}^{y_{f}}dy\ \left(1+\frac{\tilde{m}_{c}^{2}}{T^{2}}\right)\delta(s-\tilde{m}_{c}^{2})\ ,
\end{eqnarray}

\begin{eqnarray}
\rho_{8}^{3}(s)&=&-\frac{m_{c}^{2}\left(\langle\bar{q}q\rangle\langle\bar{s}g_{s}\sigma Gs\rangle+\langle\bar{s}s\rangle\langle\bar{q}g_{s}\sigma Gq\rangle\right)}{48\pi^{2}}\int_{y_{i}}^{y_{f}}dy\ \left(1+\frac{\tilde{m}_{c}^{2}}{T^{2}}\right)\delta(s-\tilde{m}_{c}^{2}) \nonumber\\
&&+\frac{m_{c}^{2}\left(\langle\bar{q}q\rangle\langle\bar{s}g_{s}\sigma Gs\rangle+\langle\bar{s}s\rangle\langle\bar{q}g_{s}\sigma Gq\rangle\right)}{96\pi^{2}}\int_{y_{i}}^{y_{f}}dy\ \frac{1}{y(1-y)}\delta(s-\tilde{m}_{c}^{2}) \nonumber\\
&&-\frac{m_{s}m_{c}\left(2\langle\bar{q}q\rangle\langle\bar{s}g_{s}\sigma Gs\rangle+3\langle\bar{s}s\rangle\langle\bar{q}g_{s}\sigma Gq\rangle\right)}{288\pi^{2}}\int_{y_{i}}^{y_{f}}dy\ \left(1+\frac{\tilde{m}_{c}^{2}}{T^{2}}+\frac{\tilde{m}_{c}^{4}}{2T^{4}}\right)\delta(s-\tilde{m}_{c}^{2}) \nonumber\\
&&+\frac{m_{s}m_{c}\langle\bar{s}s\rangle\langle\bar{q}g_{s}\sigma Gq\rangle}{192\pi^{2}}\int_{y_{i}}^{y_{f}}dy\ \left\{\frac{(1-y)}{y}+\frac{y}{(1-y)}\right\}\left(1+\frac{\tilde{m}_{c}^{2}}{T^{2}}\right)\delta(s-\tilde{m}_{c}^{2})\ ,
\end{eqnarray}

\begin{eqnarray}
\rho_{8}^{4}(s)&=&-\frac{m_{c}^{2}\left(\langle\bar{q}q\rangle\langle\bar{s}g_{s}\sigma Gs\rangle+\langle\bar{s}s\rangle\langle\bar{q}g_{s}\sigma Gq\rangle\right)}{12\pi^{2}}\int_{y_{i}}^{y_{f}}dy\ \left(1+\frac{\tilde{m}_{c}^{2}}{T^{2}}\right)\delta(s-\tilde{m}_{c}^{2}) \nonumber\\
&&+\frac{m_{c}^{2}\left(\langle\bar{q}q\rangle\langle\bar{s}g_{s}\sigma Gs\rangle+\langle\bar{s}s\rangle\langle\bar{q}g_{s}\sigma Gq\rangle\right)}{72\pi^{2}}\int_{y_{i}}^{y_{f}}dy\ \frac{1}{y(1-y)}\delta(s-\tilde{m}_{c}^{2}) \nonumber\\
&&-\frac{m_{s}m_{c}\left(2\langle\bar{q}q\rangle\langle\bar{s}g_{s}\sigma Gs\rangle+3\langle\bar{s}s\rangle\langle\bar{q}g_{s}\sigma Gq\rangle\right)}{144\pi^{2}}\int_{y_{i}}^{y_{f}}dy\ \left(1+\frac{\tilde{m}_{c}^{2}}{T^{2}}+\frac{\tilde{m}_{c}^{4}}{2T^{4}}\right)\delta(s-\tilde{m}_{c}^{2}) \nonumber\\
&&+\frac{m_{s}m_{c}\langle\bar{s}s\rangle\langle\bar{q}g_{s}\sigma Gq\rangle}{144\pi^{2}}\int_{y_{i}}^{y_{f}}dy\ \left(1+\frac{\tilde{m}_{c}^{2}}{T^{2}}\right)\delta(s-\tilde{m}_{c}^{2})\ ,
\end{eqnarray}

\begin{eqnarray}
\rho_{10}^{3}(s)&=&\frac{m_{c}^{2}\langle\bar{q}g_{s}\sigma Gq\rangle\langle\bar{s}g_{s}\sigma Gs\rangle}{192\pi^{2}T^{6}}\int_{y_{i}}^{y_{f}}dy\ \tilde{m}_{c}^{4}\delta(s-\tilde{m}_{c}^{2}) \nonumber\\
&&-\frac{m_{c}^{4}\langle\bar{q}q\rangle\langle\bar{s}s\rangle}{216T^{4}}\langle\frac{\alpha_{s}GG}{\pi}\rangle\int_{y_{i}}^{y_{f}}dy\ \left\{\frac{1}{y^{3}}+\frac{1}{(1-y)^{3}}\right\}\delta(s-\tilde{m}_{c}^{2}) \nonumber\\
&&+\frac{m_{c}^{2}\langle\bar{q}q\rangle\langle\bar{s}s\rangle}{72T^{2}}\langle\frac{\alpha_{s}GG}{\pi}\rangle\int_{y_{i}}^{y_{f}}dy\ \left\{\frac{1}{y^{2}}+\frac{1}{(1-y)^{2}}\right\}\delta(s-\tilde{m}_{c}^{2}) \nonumber\\
&&-\frac{m_{c}^{2}\langle\bar{q}g_{s}\sigma Gq\rangle\langle\bar{s}g_{s}\sigma Gs\rangle}{192\pi^{2}T^{4}}\int_{y_{i}}^{y_{f}}dy\ \frac{1}{y(1-y)}\tilde{m}_{c}^{2}\delta(s-\tilde{m}_{c}^{2}) \nonumber\\
&&+\frac{m_{c}^{2}\langle\bar{q}g_{s}\sigma Gq\rangle\langle\bar{s}g_{s}\sigma Gs\rangle}{128\pi^{2}T^{2}}\int_{y_{i}}^{y_{f}}dy\ \frac{1}{y(1-y)}\delta(s-\tilde{m}_{c}^{2}) \nonumber\\
&&+\frac{m_{c}^{2}\langle\bar{q}q\rangle\langle\bar{s}s\rangle}{216T^{6}}\langle\frac{\alpha_{s}GG}{\pi}\rangle\int_{y_{i}}^{y_{f}}dy\ \tilde{m}_{c}^{4}\delta(s-\tilde{m}_{c}^{2}) \nonumber\\
&&+\frac{m_{s}m_{c}\langle\bar{q}g_{s}\sigma Gq\rangle\langle\bar{s}g_{s}\sigma Gs\rangle}{1152\pi^{2}T^{8}}\int_{y_{i}}^{y_{f}}dy\ \tilde{m}_{c}^{6}\delta(s-\tilde{m}_{c}^{2}) \nonumber\\
&&+\frac{m_{s}m_{c}^{3}\langle\bar{q}q\rangle\langle\bar{s}s\rangle}{864T^{4}}\langle\frac{\alpha_{s}GG}{\pi}\rangle\int_{y_{i}}^{y_{f}}dy\ \left\{\frac{1}{y^{3}}+\frac{1}{(1-y)^{3}}\right\} \left(1-\frac{\tilde{m}_{c}^{2}}{T^{2}}\right)\delta(s-\tilde{m}_{c}^{2}) \nonumber\\
&&+\frac{m_{s}m_{c}\langle\bar{q}q\rangle\langle\bar{s}s\rangle}{288T^{4}}\langle\frac{\alpha_{s}GG}{\pi}\rangle\int_{y_{i}}^{y_{f}}dy\ \left\{\frac{(1-y)}{y^{2}}+\frac{y}{(1-y)^{2}}\right\}\tilde{m}_{c}^{2}\delta(s-\tilde{m}_{c}^{2}) \nonumber\\
&&-\frac{m_{s}m_{c}\langle\bar{q}g_{s}\sigma Gq\rangle\langle\bar{s}g_{s}\sigma Gs\rangle}{1152\pi^{2}T^{6}}\int_{y_{i}}^{y_{f}}dy\ \left\{\frac{(1-y)}{y}+\frac{y}{(1-y)}\right\}\tilde{m}_{c}^{4}\delta(s-\tilde{m}_{c}^{2}) \nonumber\\
&&+\frac{m_{s}m_{c}\langle\bar{q}q\rangle\langle\bar{s}s\rangle}{1728T^{8}}\langle\frac{\alpha_{s}GG}{\pi}\rangle\int_{y_{i}}^{y_{f}}dy\ \tilde{m}_{c}^{6}\delta(s-\tilde{m}_{c}^{2})\ ,
\end{eqnarray}

\begin{eqnarray}
\rho_{10}^{4}(s)&=&\frac{m_{c}^{2}\langle\bar{q}g_{s}\sigma Gq\rangle\langle\bar{s}g_{s}\sigma Gs\rangle}{48\pi^{2}T^{6}}\int_{y_{i}}^{y_{f}}dy\ \tilde{m}_{c}^{4}\delta(s-\tilde{m}_{c}^{2}) \nonumber\\
&&-\frac{m_{c}^{4}\langle\bar{q}q\rangle\langle\bar{s}s\rangle}{54T^{4}}\langle\frac{\alpha_{s}GG}{\pi}\rangle\int_{y_{i}}^{y_{f}}dy\ \left\{\frac{1}{y^{3}}+\frac{1}{(1-y)^{3}}\right\}\delta(s-\tilde{m}_{c}^{2}) \nonumber\\
&&+\frac{m_{c}^{2}\langle\bar{q}q\rangle\langle\bar{s}s\rangle}{18T^{2}}\langle\frac{\alpha_{s}GG}{\pi}\rangle\int_{y_{i}}^{y_{f}}dy\ \left\{\frac{1}{y^{2}}+\frac{1}{(1-y)^{2}}\right\}\delta(s-\tilde{m}_{c}^{2}) \nonumber\\
&&+\frac{m_{c}^{2}\langle\bar{q}q\rangle\langle\bar{s}s\rangle}{54T^{2}}\langle\frac{\alpha_{s}GG}{\pi}\rangle\int_{y_{i}}^{y_{f}}dy\ \frac{1}{y(1-y)}\delta(s-\tilde{m}_{c}^{2}) \nonumber\\
&&-\frac{m_{c}^{2}\langle\bar{q}g_{s}\sigma Gq\rangle\langle\bar{s}g_{s}\sigma Gs\rangle}{144\pi^{2}T^{4}}\int_{y_{i}}^{y_{f}}dy\ \frac{1}{y(1-y)}\tilde{m}_{c}^{2}\delta(s-\tilde{m}_{c}^{2}) \nonumber\\
&&+\frac{m_{c}^{2}\langle\bar{q}g_{s}\sigma Gq\rangle\langle\bar{s}g_{s}\sigma Gs\rangle}{32\pi^{2}T^{2}}\int_{y_{i}}^{y_{f}}dy\ \frac{1}{y(1-y)}\delta(s-\tilde{m}_{c}^{2}) \nonumber\\
&&+\frac{m_{c}^{2}\langle\bar{q}q\rangle\langle\bar{s}s\rangle}{54T^{6}}\langle\frac{\alpha_{s}GG}{\pi}\rangle\int_{y_{i}}^{y_{f}}dy\ \tilde{m}_{c}^{4}\delta(s-\tilde{m}_{c}^{2}) \nonumber\\
&&+\frac{m_{s}m_{c}\langle\bar{q}g_{s}\sigma Gq\rangle\langle\bar{s}g_{s}\sigma Gs\rangle}{576\pi^{2}T^{8}}\int_{y_{i}}^{y_{f}}dy\ \tilde{m}_{c}^{6}\delta(s-\tilde{m}_{c}^{2}) \nonumber\\
&&+\frac{m_{s}m_{c}\langle\bar{q}q\rangle\langle\bar{s}s\rangle}{144T^{4}}\langle\frac{\alpha_{s}GG}{\pi}\rangle\int_{y_{i}}^{y_{f}}dy\ \left\{\frac{1}{y^{2}}+\frac{1}{(1-y)^{2}}\right\}\tilde{m}_{c}^{2}\delta(s-\tilde{m}_{c}^{2}) \nonumber\\
&&+\frac{m_{s}m_{c}^{3}\langle\bar{q}q\rangle\langle\bar{s}s\rangle}{432T^{4}}\langle\frac{\alpha_{s}GG}{\pi}\rangle\int_{y_{i}}^{y_{f}}dy\ \left\{\frac{1}{y^{3}}+\frac{1}{(1-y)^{3}}\right\}\left(1-\frac{\tilde{m}_{c}^{2}}{T^{2}}\right)\delta(s-\tilde{m}_{c}^{2}) \nonumber\\
&&-\frac{m_{s}m_{c}\langle\bar{q}q\rangle\langle\bar{s}s\rangle}{216T^{4}}\langle\frac{\alpha_{s}GG}{\pi}\rangle\int_{y_{i}}^{y_{f}}dy\ \frac{1}{y(1-y)}\tilde{m}_{c}^{2}\delta(s-\tilde{m}_{c}^{2}) \nonumber\\
&&-\frac{m_{s}m_{c}\langle\bar{q}g_{s}\sigma Gq\rangle\langle\bar{s}g_{s}\sigma Gs\rangle}{864\pi^{2}T^{6}}\int_{y_{i}}^{y_{f}}dy\ \tilde{m}_{c}^{4}\delta(s-\tilde{m}_{c}^{2}) \nonumber\\
&&+\frac{m_{s}m_{c}\langle\bar{q}q\rangle\langle\bar{s}s\rangle}{864T^{8}}\langle\frac{\alpha_{s}GG}{\pi}\rangle\int_{y_{i}}^{y_{f}}dy\ \tilde{m}_{c}^{6}\delta(s-\tilde{m}_{c}^{2})\ ,
\end{eqnarray}
where $y_{f}=\frac{1+\sqrt{1-4m_{c}^{2}/s}}{2}$, $y_{i}=\frac{1-\sqrt{1-4m_{c}^{2}/s}}{2}$, $z_{i}=\frac{ym_{c}^{2}}{ys-m_{c}^{2}}$, $\hat{m}_{c}^{2}=\frac{(y+z)m_{c}^{2}}{yz}$, $\tilde{m}_{c}^{2}=\frac{m_{c}^{2}}{y(1-y)}$, $\int_{y_{i}}^{y_{f}}dy\rightarrow\int_{0}^{1}$, $\int_{z_{i}}^{1-y}dz\rightarrow\int_{0}^{1-y}dz$, when the $\delta$ functions $\delta(s-\hat{m}_{c}^{2})$ and $\delta(s-\tilde{m}_{c}^{2})$ appear.

\end{document}